\documentclass{article}
\usepackage{forest}
\usepackage{amsmath}
\usepackage{amssymb}
\usepackage[a4paper]{geometry}
\usepackage{graphicx}
\usepackage[T1]{fontenc}
\usepackage{mathtools}
\usepackage{thmtools}
\usepackage{xcolor}
\usepackage{nameref}
\usepackage{lipsum}
\usepackage{bbm}
\usepackage[utf8]{inputenc}
\usepackage{float}
\usepackage{thmtools}
\usepackage{xcolor}
\usepackage{nameref}
\usepackage{dsfont}
\usepackage{times}
\usepackage{multirow}
\usepackage{amssymb}
\usepackage{etoolbox}
\usepackage{natbib}
\usepackage{placeins}
\usepackage{subfigure}
\usepackage{enumitem}  
\usepackage{changepage}  
\usepackage{url}
\usepackage{xr}
\newtheorem{theorem}{Theorem}

\newtheorem{corollary}{Corollary}
\newtheorem{remark}{Remark}
\newtheorem{definition}{Definition}
\newtheorem{procedure}{Procedure}
\makeatletter
\patchcmd{\@makecaption}
{\parbox}
{\advance\@tempdima-\fontdimen2} 
{}{}
\makeatother  

\DeclareMathOperator{\Min}{Min}

\begin{document}

\title{Setwise Hierarchical Variable Selection and the Generalized Linear Step-Up Procedure for False Discovery Rate Control}

\author{Sarah Organ*, Toby Kenney*, Hong Gu\\
 * Equal co-authors \\
	Department of Mathematics and Statistics, Dalhousie University, Halifax NS, Canada}
\maketitle

\begin{abstract}

Controlling the false discovery rate (FDR) in variable selection
becomes challenging when predictors are correlated, as existing
methods often exclude all members of correlated groups and
consequently perform poorly for prediction. We introduce a new setwise
variable‑selection framework that identifies clusters of potential
predictors rather than forcing selection of a single variable. By
allowing any member of a selected set to serve as a surrogate
predictor, our approach supports strong predictive performance while
maintaining rigorous FDR control.  We construct sets via hierarchical
clustering of predictors based on correlation, then test whether each
set contains any non‑null effects. Similar clustering and setwise
selection have been applied in the familywise error rate (FWER)
control regime, but previous research has been unable to overcome the
inherent challenges of extending this to the FDR control framework.

To control the FDR, we develop substantial generalizations of linear
step‑up procedures, extending the Benjamini–Hochberg and
Benjamini–Yekutieli methods to accommodate the logical dependencies
among these composite hypotheses. We prove that these procedures
control the FDR at the nominal level and highlight their broader
applicability.  Simulation studies and real‑data analyses show that
our methods achieve higher power than existing approaches while
preserving FDR control, yielding more informative variable selections
and improved predictive models.
\end{abstract}
\subsection*{keywords}FDR control, multiple hypothesis testing, linear step-up procedures, hierarchical clustering, variable selection

\section{Introduction}

\subsection{Background}

Variable selection is a broad and central problem in statistics, vital
both for scientific interpretation and for building effective
predictive models. Reflecting its importance, a wide range of methods
have been developed. These methods fall into two broad categories:
approaches aimed at optimizing prediction and approaches designed to
control the false discovery rate (FDR).

Among prediction-oriented methods, shrinkage techniques play a
dominant role, most notably the LASSO~\citep{lasso} and its many
extensions, which incorporate additional
structure~\citep{group_lasso}, improve theoretical
properties~\citep{adap_lasso}, or enhance sparsity and
stability~\citep{Stability,SuRF}. Other prominent approaches are based
on search strategies, such as Best Subset
Selection~\citep{MIO,BestSubset}.

Some methods for controlling the false discovery rate (FDR) or
familywise error rate (FWER) rely on hypothesis testing procedures
such as BH~\citep{BH} and BY~\citep{BY}. More recent approaches
include the mirror method~\citep{gaussianmirror} with data splitting
(DS) or multiple data splitting (MDS)~\citep{mirrors}, and the model-X
knockoff framework~\citep{knockoff1}. These methods generate
artificial null statistics and compare them with observed test
statistics to identify significant predictors. While this enables
application in high-dimensional settings, it restricts the methods to
specific model families and requires knowledge of the joint
distribution of all features --- information that is typically
unavailable and difficult to estimate. Misspecification of this
distribution can lead to substantial power loss~\citep{DS} or inflated
FDR~\citep{knockoff2}.

At first glance, the distinction between prediction-focused methods
and those aimed at controlling false discoveries may seem puzzling:
true variables tend to be strong predictors, whereas including false
variables typically harms prediction. We might therefore expect that
procedures designed to select true predictors while avoiding false
ones would also yield strong predictive performance. However, when two
or more predictors are highly correlated, it becomes difficult to
identify which predictor carries the true signal, even when we are
confident that at least one does. Methods that control the FDR often
respond by selecting neither variable. This loss of power and the
potential for inflated FDR in the presence of surrogate variables have
been documented repeatedly, both for linear step-up procedures based
on $p$-values~\citep{correlation} and for more recent mirror-based
methods~\citep{DS,highcor}.

To remedy this limitation, we consider a different paradigm, where,
in addition to selecting individual variables that we are confident
are true predictors, we also select sets of predictors that we are
confident contain at least one true predictor, even if the true
predictor cannot be identified within the set. For prediction
purposes, any predictor from such a set can serve as an effective
predictor. As we do not claim to identify the true variable
within the set, a false discovery occurs only if none of the variables
in the selected set is a true predictor. This formulation preserves
FDR control while providing useful information about the location of
the true signals.

This paradigm has previously been applied to FWER controlled variable
selection. In particular, \citet{Meinshausen2008} develops an
FWER-controlling procedure which tests clusters of predictors from a
correlation-based hierarchical clustering, allowing for the selection
of clusters of predictors when the true predictor(s) cannot be
precisely determined. The tests are conducted sequentially from the
largest clusters to the smallest, with a cluster (which might be a
single predictor) only tested if all superset clusters have already
been rejected. \citet{GoemanFinos2012}, simultaneously improve and
generalise this method. The key idea of \citet{GoemanFinos2012} can be
seen as an extension of the Bonferroni-Holm method~\citep{Holm1979},
which dynamically adjusts the rejection thresholds with each
rejection, to a more general ``inheritance principle'' where the
adjustment is based on the structure underlying the hypotheses.

This line of research is related to closed testing
procedures~\citep{Marcus} for controlling FWER. These procedures test
all intersections of null hypotheses, starting with the strongest null
hypothesis, and only test weaker hypotheses if all stronger hypotheses
have already been rejected. This is equivalent to replacing the
$p$-value of each hypothesis by the largest $p$-value among all
stronger hypotheses. \citet{Hommel} shows that rejecting all
intersection hypotheses with adjusted $p$-values below $q$, controls
the FWER at level $q$.

Applied to variable selection, the null hypothesis for a set of
predictors is the intersection of the null hypotheses for its
individual predictors. Under closed testing, a set of predictors would
be selected only if the hypotheses for all supersets could also be
rejected at the threshold $q$. However, performing separate tests for
all possible subsets of predictors quickly becomes computationally
infeasible. Therefore, closed testing is often restricted to settings
where $p$-values for intersection hypotheses are computed directly
from the $p$-values of individual hypotheses, which is not appropriate
for the variable selection context.

Athough the looser FDR control is often preferred for variable
selection over FWER control, setwise variable selection with FDR
control does not appear in the literature. The major technical
obstruction, which we overcome in this paper, is appropriately
defining the number of discoveries. This is essential, as an
appropriate way to count discoveries is built into the definition of
FDR control, unlike FWER control, which is defined in terms of a
binary event (the existence of any false discoveries).  Once we have
defined an appropriate FDR for our context, we need to build a
generalised linear step-up procedure capable of handling this FDR.

While there is, to the best of our knowledge, no existing FDR-control
procedure in the literature that is suitable for setwise variable
selection, some developments in FDR control do represent important
steps in this direction. In particular,~\citet{Yekutieli2006} give
three definitions for FDR when hypotheses arranged {\it a priori} in a
hierarchical tree. While the methods developed in that paper are
limited both by the assumption of independent $p$-values, and by a
cut-off based on a worst-case tree structure often resulting in
overconservative FDR control, even under the assumed independence, the
definition of ``outer node FDR'' is an important step. The key idea is
that as rejection of weaker hypotheses makes rejection of the stronger
hypotheses redundant, we should only count the weakest rejected null
hypotheses as discoveries. However, the outer node FDR weights all
hypotheses equally, regardless of level, so that improving the
resolution of an existing discovery does not count as a discovery.
In order to build a suitable definition of FDR for selecting sets of
variables, this ``outer node FDR'' needs to be combined with weighted
FDR~\citep{wBH,HighDim}, which allows more important hypotheses to be
assigned higher weights. Other definitions of FDR for trees of
hypotheses, such as stepwise FDR for trees~\citep{treeBH}, or for
general partially ordered sets~\citep{DAGGER}, are not an appropriate
definition of FDR for variable selection.

Another prominent research direction in the field is control of
multiple different FDRs, for example the $p$-filter
method~\citep{Ramdas2019} simultaneously controls FDR for multiple
partitions of the hypotheses. Common applications of this approach are
when the partitions correspond to levels of resolution in an
hierarchical tree (e.g. taxonomic levels). This approach has been
adapted to the knockoff method~\citep{Katsevich2019,Sesia2020}. While
these methods are appropriate for some problems, the setwise
hypotheses obtained using hierarchical clustering do not naturally
fall into levels.

A common feature of the FDR and FWER control methods for hierarchical
trees described above, is that they only reject hypotheses when all
stronger null hypotheses have small $p$-values. This makes sense,
since if we reject an hypothesis, we implicitly reject any stronger
hypotheses. However, the corresponding test statistics might not
follow this pattern, so these methods can fail to reject lower-level
null hypotheses despite strong evidence against them,
resulting in lower power. We will therefore develop methods both with
and without this requirement.

\subsection{Example}

We demonstrate the pitfalls of FDR-controlled variable selection, and
the potential benefits of selecting sets of variables, on a small
simulated linear regression dataset with $n = 300$ samples and $p =
50$ predictors, where $X \sim N(0,\Sigma)$, for a covariance matrix
$\Sigma$ given by
$$\Sigma_{ij}=\left\{\begin{array}{ll}
1&\textrm{if }i=j\\
0.9 &\textrm{if }\{i,j\}=\{1,2\}\\
  0&\textrm{otherwise}
\end{array}\right.$$
Variables $X_2$ -- $X_6$ are true variables with coefficients
$\beta_i\sim N(0,1)$.


\begin{table}
  \caption{\label{example} Performance of different FDR controlled variable selection methods on example data}
  \centering
  \begin{tabular}{cccc}
    \hline
    Method & Variables Selected & FDR & MSE \\
    \hline
    BH &  $X_3$, $X_4$, $X_5$, $X_6$ & 0\% & 1.93\\	
    Mirror Method DS &  None & 0\% & NA \\
    Mirror Method MDS	& $X_3$, $X_4$, $X_5$, $X_6$ & 0\% & 1.93\\
    Knockoff &  None & 0\% & NA \\
    Setwise Selection & $\{X_1,X_2\}$, $X_3$,
    $X_4$, $X_5$, $X_6$ & 0\% & 1.72 \\
    LASSO & $X_1$, $X_2$, $X_3$, $X_4$, $X_5$, $X_6$, $X_7$, $X_{12}$, $X_{13}$, $X_{14}$, $X_{15}$, $X_{16}$, $X_{17}$ & 61.5\% & 1.69\\
    \hline
  \end{tabular}
\end{table}

Table~\ref{example} shows the results of various variable selection
methods with FDR control level $q =0.05$ on this simulated dataset,
including the method developed later in this paper, and the predictive
MSE of the selected models on a test data set of size $n = 3000$. None
of the existing FDR control methods is able to select $X_2$ because of
the high correlation with $X_1$. The LASSO method, which does not
perform FDR control, selects all true variables, and thus achieves a
good prediction despite selecting several false variables. Being
unable to distinguish which of $X_1$ and $X_2$ is a true variable, our
method selects the set $\{X_1,X_2\}$ meaning that one of them is a
true variable. This allows the method to maintain control of the FDR,
but still achieve good prediction.  While this example is contrived to
demonstrate the point, this level of correlation between predictors is
common in real data. The microbiome data example presented later in
this study is an example of this. Other examples include genetic data,
where biomarker selection with FDR control is often a crucial step,
and is especially challenging when the predictors are not only
correlated but also non-Gaussian distributed.

\subsection{Preliminaries}\label{SecPreliminaries}

We now consider the technical aspects of the approach --- for
example: which sets of predictors should we consider, how should we
select them, and how do we define FDR in this context?

We begin by setting up the exact details of our context. We consider a
general predictive modelling problem.  We have a set
$\{X_1,\ldots,X_p\}$ of predictor variables, and a response variable
$Y$. Our aim is to predict the conditional distribution of
$Y|X_1,\ldots,X_p$. Some literature defines true predictors as
predictors that are not conditionally independent of $Y$ conditional
on all other predictors. However, this definition fails in cases where
one predictor is a function of others (for example in a linear
regression with non-linear or interaction terms). Therefore, we
provide a more careful definition that extends to this context. Let
${\mathcal D}$ be the family of possible conditional distributions ---
usually $\mathcal D$ will be a parametric family, such as the family
of all normal distributions. Our model space consists of a sequence of
function spaces $({\mathcal M}_n)_{n\in \mathbb N}$, where each
${\mathcal M}_n$ is a set of functions from ${\mathbb R}^n$ to
$\mathcal D$. For example, for Gaussian linear regression, $\mathcal
D$ is the family of normal distributions with mean $\mu$ and variance
$\sigma^2$, and ${\mathcal M}_n$ consists of functions $f:{\mathbb
  R}^n\longrightarrow {\mathcal D}$ of the form
$\mu=\beta_0+\sum_{i=1}^n X_i\beta_i$, and $\sigma^2$ is constant. We
generally assume that for any $J\subseteq\{1,\ldots,n\}$, and any
$f\in{\mathcal M}_{|J|}$, the function $f_{J}$ defined by
$f_{J}(x_1,\ldots,x_n)=f((x_i)_{i\in J})$ is in ${\mathcal M}_n$,
i.e. models are nested, but this assumption is not crucial. We assume
that the true conditional distribution is given by a function of the
form $(Y|X_1=x_1,\ldots,X_p=x_p)=f((x_i)_{i\in T})$ for some unique
minimal $T\subseteq\{1,\ldots, p\}$ and $f\in{\mathcal M}_{|T|}$. A
predictor $X_i$ is a true variable if $i\in T$, and a null variable if
$i\not\in T$. For any subset $C\subseteq\{1,\ldots,p\}$, we can
perform an hypothesis test, for example a likelihood ratio test, for
the hypothesis $H_{C}$: $C\cap T=\emptyset$.

Since the problems 
arise for highly correlated variables, we perform a hierarchical
clustering of the predictors, based on correlation.  Let $\mathcal C$
denote the collection of all clusters, or subsets of $\{1,\ldots,p\}$
identified by this hierarchical clustering, including singleton
clusters. For each cluster $C\in{\mathcal C}$, we test the null
hypothesis $H_C$, using the response variable $Y$. If the test
statistic's distribution is based on the conditional distribution of
$Y$ given $X$, then the fact that we chose which subsets to test based
on the correlation of the predictors $X$ will not cause selection bias
in the resulting $p$-values, unlike screening methods based on the
response variable $Y$.

Next we consider how to define FDR in our context, and use it to build
methods that control FDR. First we recall the standard linear step-up
procedure used by the BH and BY methods. For this procedure, there are
a set $\{H_1,\ldots,H_m\}$ of null hypotheses, and corresponding
$p$-values $P_1,\ldots,P_m$, such that the $p$-values of true null
hypotheses are marginally uniformly distributed. An FDR control method
is a function that for each vector of $p$-values, selects a subset $R$
of the hypotheses to reject. If $I_N$ is the set of true null
hypotheses, then the FDR is defined as
$$\textrm{FDR}={\mathbb E}\left(\frac{|R\cap I_N|}{|R|}\right)$$ The
linear step-up method was originally expressed as ranking the
$p$-values $P_{(1)}\leqslant P_{(2)}\leqslant\cdots\leqslant P_{(m)}$,
then rejecting the smallest $k$ $p$-values, where $k$ is the largest
rank, such that $P_{(k)}\leqslant\frac{k}{\alpha}$, where
$\alpha=\frac{m}{q}$ for the BH method, and
$\alpha=\frac{m\sum_{i=1}^m\frac{1}{i}}{q}$ for the BY method. The
procedure can be rewritten as ``Reject all $p$-values less than or
equal to $c$, where $c$ is the largest cut-off satisfying
$|\{i|P_i\leqslant c\}|\geqslant \alpha c$.'' This formulation, used
by~\citet{finnerBH}, to develop a nonlinear threshold function in
place of the linear $\alpha c$, is easier to generalize to our conext.

The FDR can be thought of as $${\mathbb E}\left(\frac{\textrm{number
    of false discoveries}}{\textrm{total number of
    discoveries}}\right)$$ so to define it in our context, we just
need to determine how to count the number of discoveries. The issue is
that if we, for example, test the null hypotheses:

\begin{description}
  \item[$H_{\{1\}}$:] The variable $X_1$ is not a true predictor.

  \item[$H_{\{2\}}$:] The variable $X_2$ is not a true predictor.

  \item[$H_{\{1,2\}}$:] Neither of the variables $X_1$ nor $X_2$ is a
    true predictor.

\end{description}

We cannot legitimately count rejecting all three of these hypotheses
to represent three discoveries, as rejecting all three hypotheses
corresponds to selecting only two variables. Additionally, since
$H_{\{1,2\}}$ here is a stronger null hypothesis, and therefore easier to
reject, we should consider its rejection to be less of a discovery
than rejecting either $H_{\{1\}}$ or $H_{\{2\}}$.

For our setwise variable selection situation, we adopt the following
approach to counting discoveries. We first assign a
weight $\phi_C\geqslant 0$ to each hypothesis $H_C$, based on the
number of variables in $C$. In this paper, we use
$\phi_C=\frac{1}{|C|}$, so that selecting a set of $k$ variables
corresponds to $\frac{1}{k}$ discoveries, though our method works for
more general $\phi_C$. This has the good feature that if $C$ is a
singleton, $\phi_C=1$, so that selecting one variable counts as a
single discovery.

If there is an implication between two rejected null hypotheses, we
only count the weakest rejected hypothesis as a discovery, since after
rejecting the weaker hypothesis, rejecting the stronger hypothesis is
no longer a discovery. (\citet{Yekutieli2006} refer to this, for
unweighted hypotheses, as ``outer nodes FDR''.) Thus, our number of
(false) discoveries is given by
 $$\sigma({\mathcal S})=\sum_{C\in\Min({\mathcal S})} \phi_C$$ where
${\mathcal S}\subseteq{\mathcal C}$ is the set of (falsely) rejected
clusters, and $\Min({\mathcal S})$ is the set of minimal clusters in
$\mathcal S$ (i.e. clusters that do not contain any other clusters in
$\mathcal S$). Because of the hierarchical nature of the clustering
$\mathcal C$, for any ${\mathcal S}\subseteq{\mathcal C}$, the minimal
clusters in $\mathcal S$ will be disjoint. We will refer to the
function $\sigma$ that counts the number of discoveries as a ``sizing
function''.

Another issue we need to consider is that because of the implications
between hypotheses, it makes no sence to reject a weaker $H_{C_1}$ without also rejecting a stronger $H_{C_2}$ which
implies $H_{C_1}$. Thus, we impose a constraint on the combinations of
hypotheses that can be rejected. That is, we are given, {\it a
  priori}, a set $\mathcal R$ of sets of hypotheses that can be
rejected. The set $\mathcal R$ is closed under intersections, so for
any set $J$ of hypotheses, there is a smallest
$\overline{J}\in{\mathcal R}$ such that $J\subseteq \overline{J}$,
called the ``closure'' of $J$. We develop two procedures to ensure the
set of rejected hypotheses falls in $\mathcal R$.  Both procedures
require extensions to the existing theory of FDR control methods.


This general FDR control framework incorporates the
original BH and BY methods, and the weighted BH method (see
Section~\ref{GLSUP_general}). This level of generality has a wide
range of potential applications. For example, using a
general sizing function similar to the one we develop here may be more
suitable for applications in microbiome data analysis and GWAS than 
the sizing functions implicitly used in the hierarchical
FDR control methods of~\cite{Yekutieli2006}.


\subsection{Contributions and Overview}

Our paper includes two major contributions. The first is a broad
generalisation of FDR control methods for multiple hypothesis testing,
to incorporate cases where the ``number of discoveries'' is measured
by a general ``sizing function'', and where relations between the
hypotheses impose constraints on the rejectable sets of hypotheses. In
addition to including the original BH and BY procedures and the
weighted BH procedure~\citep{wBH}, as special cases, and providing a
solution to our problem of setwise variable selection with FDR
control, this framework has potential applications to a range of other
multiple hypothesis testing problems.

The second major contribution is a novel setwise variable selection
framework which overcomes the challenges of maintaining FDR control
while achieving adequate power in high-correlation settings. Previous
research in variable selection has been divided between methods that
aim to control FDR and methods that aim to build good predictive
models. Our setwise variable selection methods have the potential to
achieve both objectives.

The remainder of the paper is structured as follows. In
Section~\ref{SecGLSUP}, we develop the generalised linear step-up
framework for controlling FDR for a general sizing function. We then
derive theoretical results for the appropriate threshold slope to
control FDR under different assumptions about dependency between the
$p$-values. In Section~\ref{SecSHRED}, we use the generalized linear
step-up procedure to allow our Setwise Hierarchical Rate of Erroneous
Discovery (SHRED) methods to select sets of variables in a way that
controls the FDR. This approach is applicable to any variable
selection problem for which valid $p$-values are available for any set
of predictors. In Section~\ref{SecSim}, we compare our SHRED methods
with other state-of-the-art variable selection methods over a range of
simulation studies. Finally, in Section~\ref{SecReal}, we apply these
variable selection methods on a real microbiome data set measuring
taxonomic abundances in wastewater treatment
plants~\citep{midas}. Section~\ref{SecConc} concludes the paper.

\section{The Generalized Linear Step-Up Procedure (GLSUP)}\label{SecGLSUP}

\subsection{General Procedure}\label{GLSUP_general}

For our generalized linear step-up procedure, we have the
following input:

\begin{definition}\label{GLSUPData}
  The input needed to apply the GLSUP method consist of the following:
\begin{itemize}

\item We have a set $H_1,\ldots,H_m$ of null hypotheses with
  corresponding conservative $p$-values $P_1,\ldots,P_m$, which are
  random variables. (By conservitive $p$-values, we mean that under
  the null hypothesis $H_i$, $P_i$ has a superuniform distribution.)
  We let $\mathcal{P}$ denote the powerset of $\{1,\ldots,m\}$,
  i.e. $\mathcal{P} = \{J \mid J \subseteq \{1,\ldots,m\}\}$.

\item We have a given collection $\mathcal{R}\subseteq{\mathcal P}$ of
  ``rejectable'' subsets of $\{1,\ldots,m\}$, that is, sets of
  hypotheses that can be consistently rejected.  We assume ${\mathcal
    R}$ is closed under arbitrary intersections. This means that there
  is a corresponding closure operator that sends any $J\subseteq
  \{1,\ldots,m\}$ to the smallest set $\overline{J}\in{\mathcal R}$
  that contains $J$.

\item We have a sizing function $\sigma:{\mathcal
  R}\longrightarrow{\mathbb R}_{\geqslant 0}$, which is (not
  necessarily strictly) increasing with respect to set inclusion.

\end{itemize}
\end{definition}

In this framework, the generalized linear step-up method with slope
$\alpha$ is as follows:

\begin{procedure}[GLSUP]\label{GLSUP}
\ 
  
\begin{itemize}
  \item For any cut-off $c\in[0,1]$, we define $I_c=\{i|P_i \leqslant  c\}$.
    
  \item Define $c_{\textrm{max}}=\sup\left\{c\in[0,1]\middle|\sigma\!\!\left(\overline{I_c}\right)\geqslant
    \alpha c\right\}$
    
  \item Reject all hypotheses in
    $\overline{I_{c_{\textrm{max}}}}$.
\end{itemize}

\end{procedure}

This is called a linear step-up procedure because the threshold
$\alpha c$ is a linear function of the cut-off $c$. We could instead
develop a procedure with a non-linear threshold, similar
to~\citet{finnerBH}; however, that is beyond the scope of the current
paper. The slope $\alpha$ of the linear threshold is chosen to control
the FDR, and depends on our assumptions about null $p$-values. We
derive values of $\alpha$ for various common variable selection cases
in Sections~\ref{SecTheory} and~\ref{SecSHRED}.

In addition to the above notation, we let $I_N \subseteq
\{1,\ldots,m\}$ denote the indices of true null hypotheses, and $I_A =
\{1,\ldots,m\} \setminus I_N$ denote the indices of false null
hypotheses.
The generalized false discovery
rate (gFDR) and generalized power (gPower) are defined as,
\begin{equation*}
	\textrm{gFDR} = \mathbb{E} \left( \frac{\sigma\left(\overline{I_{c_{\textrm{max}}}} \cap I_N\right)}{\sigma\left(\overline{I_{c_{\textrm{max}}}}\right)} \right)
\end{equation*} 
and,
\begin{equation*}
	\textrm{gPower} =  \mathbb{E} \left( \frac{\sigma\left(\overline{I_{c_{\textrm{max}}}} \cap I_A\right)}{\sigma\left(\overline{I_A}\right)} \right)
\end{equation*}

Existing linear step-up FDR control procedures are special cases of
the GLSUP, with particular sizing functions and suitably chosen
$\alpha$ values. When ${\mathcal R}={\mathcal P}$ and $\sigma(J)=|J|$,
setting $\alpha =\frac{m}{q}$ gives the BH procedure, while $\alpha
=\frac{m \sum_{j=1}^{m} \frac{1}{j}}{q}$ gives the BY procedure, where
$q$ is the desired FDR control level. The weighted BH (wBH)
procedure~\citep{wBH} is also a special case of GLSUP with
$\sigma(J)=\sum_{i \in J} w_i$ and $\alpha
=\frac{\sum_{i=1}^mw_i}{q}$. It is easy to see that if
$\sigma(J)=|J|$ and ${\mathcal R}={\mathcal P}$, the gFDR and gPower
under the BH and BY procedures are equivalent to the standard FDR and
power, while if $\sigma(J)=\sum_{i\in J}w_i$ and ${\mathcal
  R}={\mathcal P}$, gFDR and gPower are equivalent to weighted FDR and
weighted power from~\citet{wBH}.

\subsection{Modified GLSUP Procedure}

In the variable selection problem that is our focus, rejectable sets
of hypotheses are based on implication. Reverse implication
$H_i\Leftarrow H_j$ defines a partial order on the set
$\{H_1,\ldots,H_m\}$. Rejectable sets are upward closed sets for this
partial order. In this case, we can modify our GLSUP procedure so that
the set of $p$-values less than a cut-off $c$ is always rejectable.

\begin{procedure}[mGLSUP]\label{mGLSUP}
  \

  \begin{itemize}

    \item Define
      $P_1',\ldots,P_m'$ by $P_{i}' = \sup\{P_j, j \in \{1,\ldots,m\}|(H_j
      \implies H_i)\}$

    \item For any cut-off $c\in[0,1]$, define $I_c=\{i|P'_i \leqslant  c\}$.
  
    \item Define $c_{\textrm{max}}=\sup\left\{c\in[0,1]\middle|\sigma\left(\overline{I_c}\right)\geqslant
      \alpha c\right\}$
    
    \item Reject all hypotheses in
      $\overline{I_{c_{\textrm{max}}}}$.

  \end{itemize}

\end{procedure}
Note that because of the way $P_i'$ are defined, we have
$I_c=\overline{I_c}$ for all $c\in[0,1]$.

This procedure has a similar effect to a number of other procedures in
the literature~\citep{Meinshausen2008,Yekutieli2006,treeBH}, where an
hypothesis can only be rejected if its parent hypothesis is
rejected. An advantage of this approach is that for our sizing
function $\sigma$, which is the sum of the weights of minimal
elements, and for our variable selection problem, where the partial
order has the structure of a bifurcating tree, we can express $\sigma$
as a sum of pairwise weights $\sigma(A)=\sum_{i,j\in A}w_{ij}$ for
some $w_{ij}\geqslant 0$. This is proved in
Appendix~\ref{App_IncExcIneq}. This allows us to prove gFDR control
under certain conditions, using a smaller cut-off $\alpha$, so in
certain cases, the modified GLSUP can achieve better gPower than the
GLSUP.

\subsection{Theory for the Generalized Linear Step-Up Procedure}\label{SecTheory}


In this section, we prove that with appropriate choices of $\alpha$,
our GLSUP and mGLSUP control the gFDR. We first prove gFDR control
without any assumptions on the $p$-values.

\begin{theorem}\label{GeneralizedBY}
For the inputs from Definition~\ref{GLSUPData}, suppose that for any
$J\subseteq\{1,\ldots,m\}$, $\sigma\left(\overline{J}\right)\leqslant \sum_{i\in
  J}w_i$ for some weights $w_1,\ldots,w_m$. Then the generalized
linear step-up procedure with cut-off
  $$\alpha=\frac{\left(\sum_{i\in I_N}w_i\right)\left(1+\log(\sigma(\{1,\ldots,m\}))\right)-\sum_{i\in
    I_N}w_i\log(\sigma_i)}{q}$$ where
$\sigma_i=\sigma(\{i\})$, controls the gFDR at level $q$.
\end{theorem}

The proof of Theorem~\ref{GeneralizedBY} is a straightforward
generalisation of the proof from~\cite{BY}, and is given in
Supplementary Appendix~\ref{App_TheoremProofs}.

\begin{remark}
In the case where there is no structure between hypotheses, so
$\sigma(J)=|J|$ is the cardinality, and all sets are rejectable
(i.e. $\mathcal{R}=\mathcal{P}$), we set $w_i=1=\sigma_i$, to get
$\alpha=\frac{m_0}{q}\left(1+\log(m)\right)$. As $m_0$ is unknown,
using $m$ as an upper bound is standard practice. Then
Theorem~\ref{GeneralizedBY} gives
$\alpha=\frac{m}{q}\left(1+\log(m)\right)$. This is very close to, but
slightly higher than the cut-off
$\alpha=\frac{m}{q}\sum_{i=1}^m\frac{1}{i}$ used by the BY
method~\citep{BY}. The difference is because in the original problem
in~\citep{BY}, the sizing function is the number of rejected
hypotheses, and so is constrained to be an integer.
\end{remark}

As for the original BH, we can control the gFDR using a
smaller cut-off $\alpha$ (and thus make the test less conservative) if
we assume that the $p$-values satisfy the positive regression
dependency on a subset (PRDS) condition on the set $I_N$ of true
nulls~\citep{BY}:

\begin{definition}[PRDS]
A set of $[0,1]$-valued random variables $\mathbf{P}=(P_1,\ldots,P_m)$
satisfies the {\em PRDS} condition on the subset
$I_N\subseteq\{1,\ldots,m\}$ if for any $i\in I_N$ and any up-closed
set $U\subseteq [0,1]^{m-1}$, $P(\mathbf{P}_{\hat{i}} \in U| P_i = x)$
is an increasing function in $x$, where $\mathbf{P}_{\hat{i}}$ is
$\mathbf{P}$ with the $i$th index removed.
\end{definition}

Under this assumption, we can prove gFDR control with a lower cut-off.

\begin{theorem}\label{PRDS}
Suppose that the conservative $p$-values $P_1,\ldots,P_m$ satisfy the
PRDS condition on $I_N$, and that the sizing function
$\sigma:{\mathcal P}\longrightarrow{}{\mathbb R}_{\geqslant 0}$ is
bounded by a sum of weights: $(\forall
J\subseteq\{1,\ldots,m\})(\sigma\left(\overline{J}\right)\leqslant \sum_{i\in
  J}w_i)$ for some weights $w_i\geqslant 0$. Then the generalized
linear step-up method with cut-off $\alpha=\frac{\sum_{i=1}^mw_i}{q}$
controls the gFDR at level $q$.
\end{theorem}

In the case where $\sigma(J)=|J|$ and $\mathcal{R}=\mathcal{P}$, this
is the same value of $\alpha$ as in the classical BH
method. Theorem~\ref{PRDS} is proved using a fairly straightforward
extension of the proof from~\cite{BY}. In the case where
$\sigma(J)=\sum_{i\in J}w_i$, this is the same bound as the weighted
BH method~\citep{wBH}. wFDR control was only originally proved for that
method in the case of independent $p$-values, so our theorem
represents an improvement to the theory, even for existing methods.

For our modified GLSUP, we cannot express
$\sigma\!\!\left(\overline{J}\right)$ as a sum of weights $\sum_{i\in
  J}w_i$, but we can express it as a sum of pairwise weights,
$\sum_{i,j\in J}w_{ij}$. To extend our result with smaller
$\alpha$ to this case, we generalize the PRDS condition to a Pairwise
PRDS condition:

\begin{definition}[PPRDS] A set of $[0,1]$-valued random variables $\mathbf{P}=(P_1,\ldots,P_m)$
satisfies the {\em PPRDS} property
  on the subset $J\subseteq\{1,\ldots,m\}$ if it satisfies the PRDS
  condition on $J$ and for any $i\ne j \in J$, and any upset $U
  \subseteq [0,1]^{m-2}$, we have
 	$$P(\mathbf{P}_{\hat{i},\hat{j}} \in U|P_i = x_i, P_j = x_j)$$
 	is an increasing function in $x_i$ and $x_j$, where
        $\mathbf{P}_{\hat{i},\hat{j}}$ is the vector $\mathbf{P}$ of
        $p$-values with $P_i$ and $P_j$ removed.
 \end{definition}

\begin{theorem}\label{PPRDS}
For the inputs from Definition~\ref{GLSUPData}, suppose the conservative
$p$-values satisfy the {\em Pairwise Positive Regression
    Dependency on a Subset} (PPRDS) condition on $I_N$ and the sizing
function is bounded by a sum of pairwise weights, i.e. $(\forall
J\subseteq\{1,\ldots,m\})(\sigma\left(\overline{J}\right)\leqslant \sum_{i,j\in
  J}w_{ij})$ where $w_{ij}\geqslant 0$. Then the generalized linear
step-up method with threshold $\alpha=\frac{\sum_{i,j\in
    I_N}w_{ij}}{q}$ controls the gFDR at level $q$.
\end{theorem} 

When PPRDS holds, Theorem~\ref{PRDS} is a special case
of Theorem~\ref{PPRDS} obtained by setting $w_{ij}=0$ for all $i\ne
j$. However, the PPRDS condition is stronger than the PRDS
condition. In fact, Theorems~\ref{PRDS} and~\ref{PPRDS} can both
be seen as special cases of a more general theorem, where $\sigma$
satisfies an inequality form of the inclusion-exclusion principle: for
any $J_1,\ldots,J_n\subseteq \{1,\ldots,m\}$, $$\sigma(J_1\cup\cdots\cup
J_n)\geqslant \sum_{S\subseteq\{1,\ldots,n\},S\ne\emptyset}
(-1)^{|S|-1}\sigma\!\left(\cap_{k\in S}J_k\right)$$ In this case, we
show in Supplementary Appendix~\ref{App_IncExcIneq} that
$\sigma$ can be expressed as a sum, $\sigma(J)=\sum_{J'\subseteq J}
w_{J'}$ for some weights $w_{J'}\geqslant 0$. To state the more
general theorem, we start by generalizing PRDS to
the setwise PRDS condition:

\begin{definition}[SPRDS] A set of random variables $\mathbf{P}=
  (P_1,\ldots,P_m)\in[0,1]^m$ satisfies the {\em Setwise Positive Regression
    Dependency on a Subset} (SPRDS) condition for a collection
  ${\mathcal J}\subseteq {\mathcal P}$ of subsets of $\{1,\ldots,m\}$
  if for any $C\in{\mathcal J}$, any $x,y\in [0,1]^{|C|}$
  with $(\forall i\in C)(x_i\leqslant y_i)$ and any upward-closed
  $U\subseteq[0,1]^{m-|C|}$, we have $$P(\mathbf{P}_{\widehat{C}}\in
  U|\mathbf{P}_C=x)\leqslant P(\mathbf{P}_{\widehat{C}}\in
  U|\mathbf{P}_C=y)$$
where $\mathbf{P}_{\widehat{C}}$ denotes
  $\mathbf{P}$ with indices from $C$ removed and $\mathbf{P}_{C}$
  denotes the restriction of $\mathbf{P}$ to $C$.
    \end{definition}

It is not hard to see that the PRDS and the PPRDS are special cases of
the SPRDS condition, taking $\mathcal S$ to be the set of singletons
or singletons and pairs respectively.

\begin{theorem}\label{SPRDS}
Let $H_1,\ldots,H_m$ be a set of null hypotheses. Let $I_N\subseteq
\{1,\ldots,m\}$ be the set of true nulls. Let the sizing function be
bounded by $\sigma\left(\overline{J}\right)\leqslant \sum_{J'\subseteq J}w_{J'}$
where $w_{J'}\geqslant 0$ are weights on subsets of $\{1,\ldots,m\}$.
Also suppose that the conservative $p$-values $\mathbf{P}$ satisfy the
SPRDS condition for the collection of subsets of $I_N$ with non-zero
weight $w_J>0$. Then the generalized linear step-up method with threshold
$\alpha=\frac{W}{q}$, where $W=\sum_{J\subseteq\{1,\ldots,m\}}w_J$,
controls the gFDR at level $q$.
\end{theorem}

In the case where only singleton subsets of $\{1,\ldots,m\}$ have
non-zero weights, this gives Theorem~\ref{PRDS}, and in the case where
only subsets of size at most 2 have non-zero weights, this reduces to
Theorem~\ref{PPRDS}. The proof of Theorem~\ref{SPRDS} is a generalized
version of the proof given in~\cite{BY}, and is in Supplemental
Appendix~\ref{App_TheoremProofs}.

\section{The SHRED Procedures for Setwise Variable Selection with FDR Control}\label{SecSHRED}

Recall our variable selection problem. We have a set
$\{X_1,\ldots,X_p\}$ of predictor variables, and a response variable
$Y$. Our model space consists of a family ${\mathcal D}$ of possible
conditional distributions, and a (usually nested) sequence of function
spaces $({\mathcal M}_n)_{n\in \mathbb N}$, where each ${\mathcal
  M}_n$ is a set of functions $f:{\mathbb R}^n\longrightarrow \mathcal
D$. We assume that the true conditional distribution is given by
$(Y|X_1=x_1,\ldots,X_p=x_p)=f((x_i)_{i\in T})$ for some unique minimal
$T\subseteq\{1,\ldots, p\}$ and $f\in{\mathcal M}_{|T|}$. A predictor
$X_i$ is a true variable if $i\in T$, and a null variable if $i\not\in
T$. For any subset $C\subseteq\{1,\ldots,p\}$, there is an hypothesis
test for the null hypothesis $H_{C}$: $C\cap T=\emptyset$. Our
objective is to select a collection of subsets of $\{1,\ldots,p\}$ in
a way that controls the gFDR.

\subsection{The SHRED Procedure}

Our Setwise Hierarchical Rate of Erroneous Discovery (SHRED) method
is
as follows:
\begin{enumerate}[label=(\arabic*)]

\item Perform correlation-based hierarchical clustering on the set
  $\{X_1,\ldots,X_p\}$ of all predictors. Let ${\mathcal C}$ denote
  the set of all clusters identified by this method.

\item Fit the full model predicting $Y$ from all variables in $X$.

\item For each cluster, $C_i\in{\mathcal C}$, refit the model with the
  predictors from $C_i$ removed, and use this to obtain a $p$-value
  for the null hypothesis $H_{C_i}$ that all predictors in $C_i$ are
  null variables. We use the likelihood ratio test, or $F$-test where
  appropriate, comparing this reduced model to the full
  model. However, any valid hypothesis test 
  could be used.

\item Define a ``sizing function'' on sets of hypotheses by
  $\sigma(J)=\sum_{C_i\in\Min(J)} \phi_i$ where $\Min(J)=\{C_i\in
  J|(\forall C_j\in J)(C_j\subseteq C_i\Rightarrow i=j)\}$ is the set
  of minimal subsets in $J$.

\item Apply GLSUP or mGLSUP to the $p$-values for these hypotheses to
  reject a set of these hypotheses, using one of the values of
  $\alpha$ defined in Section~\ref{SecThresholdSlope}.

\item The selected sets are the minimal rejected clusters from the
  GLSUP or mGLSUP method.
  
\end{enumerate}

We call the mGLSUP method with hierarchical clustering
``Setwise Hierarchical Rate of Erroneous Discovery with Descending
E-value Reduction'' (SHREDDER). (The term ``E-value''~\citep{evalues}
is an alternative framework, equivalent to conservative $p$-values.)

\subsection{Threshold Slope}~\label{SecThresholdSlope}

In this section, we apply Theorems~\ref{GeneralizedBY},~\ref{PRDS}
and~\ref{PPRDS} to select a cut-off $\alpha$ which guarantees gFDR
control for the sizing function $\sigma$ defined in
Section~\ref{SecPreliminaries}. Recall our situation: the null
hypotheses $H_1,\ldots,H_m$ are partially ordered by reverse
implication, and $\sigma$ is defined by $\sigma(J)=
\sum_{i\in\Min(J)}\phi_i$ where $\phi_i\geqslant 0$ are weights on
hypotheses and $\Min(J)$ denotes the set of minimal hypotheses in $J$
under reverse implication. (These correspond to the minimal subsets of
variables.) The weights $\phi_i$ are chosen so that $\sigma$ is an
increasing set function. (For variable selection, this just means that
if $H_i\Leftarrow H_j$, then $\phi_i\geqslant\phi_j$, but this theory
applies under more general implication structures.) Let
$I_N\subseteq\{1,\ldots,m\}$ be the set of indices of true null
hypotheses.

If we make no assumptions about the dependency between the null
$p$-values, from Theorem~\ref{GeneralizedBY} with $w_i=\phi_i$, we
prove:

\begin{corollary}\label{SHRED_ARB}
For any $q>0$, the generalized linear step-up procedure with
$$\alpha=\alpha_{\textrm{SHRED}}=\frac{\left(\sum_{i\in I_N}\phi_i\right)(1+\log(p))-\sum_{i\in I_N}\phi_i\log(\phi_i)}{q}$$
  where $p=\sigma\left(\{1,\ldots,m\}\right)$, controls the gFDR at
  level $q$.
\end{corollary}

In practice $\sum_{i\in I_N}\phi_i$ is unknown, and we use its upper
bound $\sum_{i=1}^{m}\phi_i$. For setwise
variable selection, we set $\phi_i=1$ for all singletons, so that $p$
is the total number of predictors.

If we assume that the $p$-values satisfy the PRDS condition, we can
obtain a smaller cut-off to control gFDR by applying Theorem~\ref{PRDS} with the bound $\sigma(J) \leqslant \sum_{i \in J}\phi_i$:

\begin{corollary}\label{prds}
If the $p$-values satisfy the PRDS condition on the true null
hypotheses, then, the generalized linear step-up procedure with
$$\alpha=\alpha_{\textrm{SHRED}_{\textrm{PRDS}}} = \frac{\sum_{i=1}^m
  \phi_i}{q}$$ controls the gFDR at level $q$.
\end{corollary}

It is straightforward to see that if we assign weight $\phi_i=0$ to all
sets of size greater than 1, and weight $\phi_i=1$ to all singletons,
then this becomes the BH procedure.

For the SHREDDER method, we first recall some elementary partial order
theory:

\begin{definition}
  For a partially ordered set $(X,\Leftarrow)$, an element $x\in X$
  {\em covers} an element $y\in X$ if $y\Leftarrow x$ and for any
  $z\in X$ with $y\Leftarrow z\Leftarrow x$, we have either $z=x$ or
  $z=y$. For example, if $(X,\Leftarrow)$ has a tree structure, then a
  node covers only its children. 
\end{definition}

\begin{corollary}\label{shredder}
Let $\Leftarrow$ be a partial order on the hypotheses, with the
property that any $H_i$ covers at most two $H_j$ in this partial
order. If the $p'$-values $p_1',\ldots ,p_m'$ given by $p_{i}' =
\sup\{p_j | H_j \!\!\implies\!\! H_i\}$ satisfy the PPRDS condition on the
true null hypotheses, then the mGLSUP procedure with
$\alpha_{\textrm{SHREDDER}} = \frac{\sigma({1,\ldots,m})}{q}$ controls
the gFDR at level $q$.
\end{corollary}

Corollary~\ref{shredder} is proved in Supplementary
Appendix~\ref{App_IncExcIneq}. The idea is to show that for any $J\in{\mathcal
  R}$, $\sigma(J)$ is a sum of pairwise weights, then use
Theorem~\ref{PPRDS}.

\begin{remark}
Corollary~\ref{SHRED_ARB} and Corollary~\ref{prds} also hold whenever
$\sigma(J)\leqslant \sum_{i\in\Min(J)}\phi_i$ --- that is, these
methods can apply whenever we have an upper bound on the sizing
function of this form. For Corollary~\ref{shredder}, if
$\sigma(J)\leqslant \sum_{i\in\Min(J)}\phi_i$, then the corollary
holds with $\alpha_{\textrm{SHREDDER}}$ replaced by
$\frac{\sum_{i\in\Min(\{H_j|j=1,\ldots,m\})}\phi_i}{q}$.
\end{remark}

In practice, the empirical gFDR achieved by SHRED using the PRDS
cut-off is almost always much lower than the nominal rate. It is
therefore possible to achieve a higher gPower while maintaining an
acceptable gFDR by choosing a smaller value for $\alpha$. We have
found that the heuristic $\alpha_{\textrm{SHRED}_{\textrm{h}}} =
\frac{p}{q}$ works well in practice. This approach uses the same gFDR
cut-off that we would use to select individual variables under the BH
method. While gFDR control is not guaranteed with this cut-off, for
our simulations the gFDR was always empirically controlled at level
$q$ using this cut-off. We include a small simulation study of how the
empirical FDR compares to a variety of theoretical FDR levels. Results
are in Supplementary Appendix~\ref{App_FDR_tables}.

The \texttt{SHRED} package implementing these methods is available on
\texttt{CRAN}. 




\section{Simulations}\label{SecSim}

\subsection{Simulation Design}\label{SecSimDes}
We assess the performance of our methods as compared to the BH, BY,
the mirror method with MDS, the mirror method with DS, and the
knockoff method, through simulations for Gaussian regression,
logistic regression, and Poisson regression (with log link). We
simulate data of the form $X \sim N(\mu,\Sigma)$, $\mu = (0,...,0)^T$,
$ \textrm{diag}(\Sigma) = 1$, under three different correlation
structures:

\begin{enumerate}[label=(\arabic*)]

\item Common component correlation:
$$\Sigma_{ij}=\left\{\begin{array}{ll}
1 & \textrm{if }i=j\\
\rho & \textrm{otherwise}
\end{array}\right.$$


\item Clustered correlation: randomly partition the variables into
  disjoint clusters $C_1,\ldots,C_K$ of sizes
  $|C_1|=5,|C_2|=10,|C_3|=15,\ldots$, with $C_K$ containing any
  leftover variables, so when $p=300$, $K=11$ and $|C_K|=25$, while
  when $p=200$, $K=9$ and $|C_K|=20$.
  $$\Sigma_{ij}=\left\{\begin{array}{ll}
1 & \textrm{if }i=j\\
\rho_{k} & \textrm{if }i\ne j\in C_k\\
0 & \textrm{otherwise}
\end{array}\right.$$
where each $\rho_k$ is independantly drawn from a uniform distribution
with a certain range. 

%

\item Autoregressive (AR(1))
  correlation:

  $$\Sigma_{ij}=\rho^{|i-j|}$$

\end{enumerate}

For each setting, 100 simulation replicates are used for Gaussian and
Poisson regression, and 200 for logistic regression due to its
tendancy to produce wider confidence intervals.

Covariance matrices are generated using the \texttt{simstudy v0.8.1 R}
package~\citep{simstudy}. A new correlation matrix is generated for
each replicate in each scenario. In all simulations $T = 100$ true
predictors are chosen at random. The coefficients $\beta_i$ are
simulated as i.i.d standard normal distributions, for $i=1,\ldots,
T$. Dataset sizes are fixed at $p = 300, n = 1000$ for Gaussian
regression, and $p = 200, n = 5000$ for logistic and Poisson
regression. The intercept term for the Logistic simulation is set so
that the marginal probability is 0.5 for each dataset, and Poisson
counts are generated with intercept fixed at -1, typically yielding
responses between 0 and 10.

Mirror methods follow~\cite{mirrors}: DS uses a single data split,
while MDS uses 50 data splits. Both DS and MDS results are obtained
using the R code given in~\cite{mirror_code}. Knockoff results are
generated using the \texttt{knockoff v0.3.6 R} package~\citep{knockoff_package}.

For each simulation and each replicate, we simulate a test set of the
same size for evaluating MSE and classification accuracy. For Poisson
regression, MSE is computed under the same approach as the Gaussian
case. All methods target gFDR control at $q = 0.05$.

The \texttt{R} code for all our simulations is available in the
following github repository~\url{https://github.com/s-organ/SHRED-R}.

\subsection{Results}

Figures~\ref{normalSHRED}--\ref{poissonSHRED} show the gPower,
gFDR, and either MSE or classification accuracy for each method under
each scenario. For all methods, mean squared error (MSE)
and classification accuracy were computed by first refitting a model
using only the variables selected by that method on the training data,
and then evaluating predictive performance on an independent test
dataset. For the SHRED methods, one variable was randomly chosen from
each selected set for this refitting.

LASSO results for MSE and classification accuracy are included in the
figures to facilitate predictive-performance comparisons, while LASSO
gPower and gFDR results are provided separately in
Appendix~\ref{App_SupTables}. LASSO variable selection was performed
using the cross-validated tuning parameter that minimized
cross-validated MSE or negative log-likelihood, and predictive
performance was evaluated for the model using the corresponding LASSO
coefficients.

For the SHRED methods, gPower and gFDR were computed by applying the
function $\sigma$ to the correctly rejected and falsely rejected null
hypotheses, respectively. For the logistic and Poisson simulations,
the knockoff method is omitted from the figures due to its
low power; similarly, the mirror method with DS is not shown in
several figures due to low performance. Full results for these methods
and the simulation tables for each figure are given in
Appendix~\ref{App_SupTables}.

\subsubsection{Gaussian Regression}

\begin{figure}[htbp]
	\centering
	\includegraphics[width=13cm, height = 10.5cm]{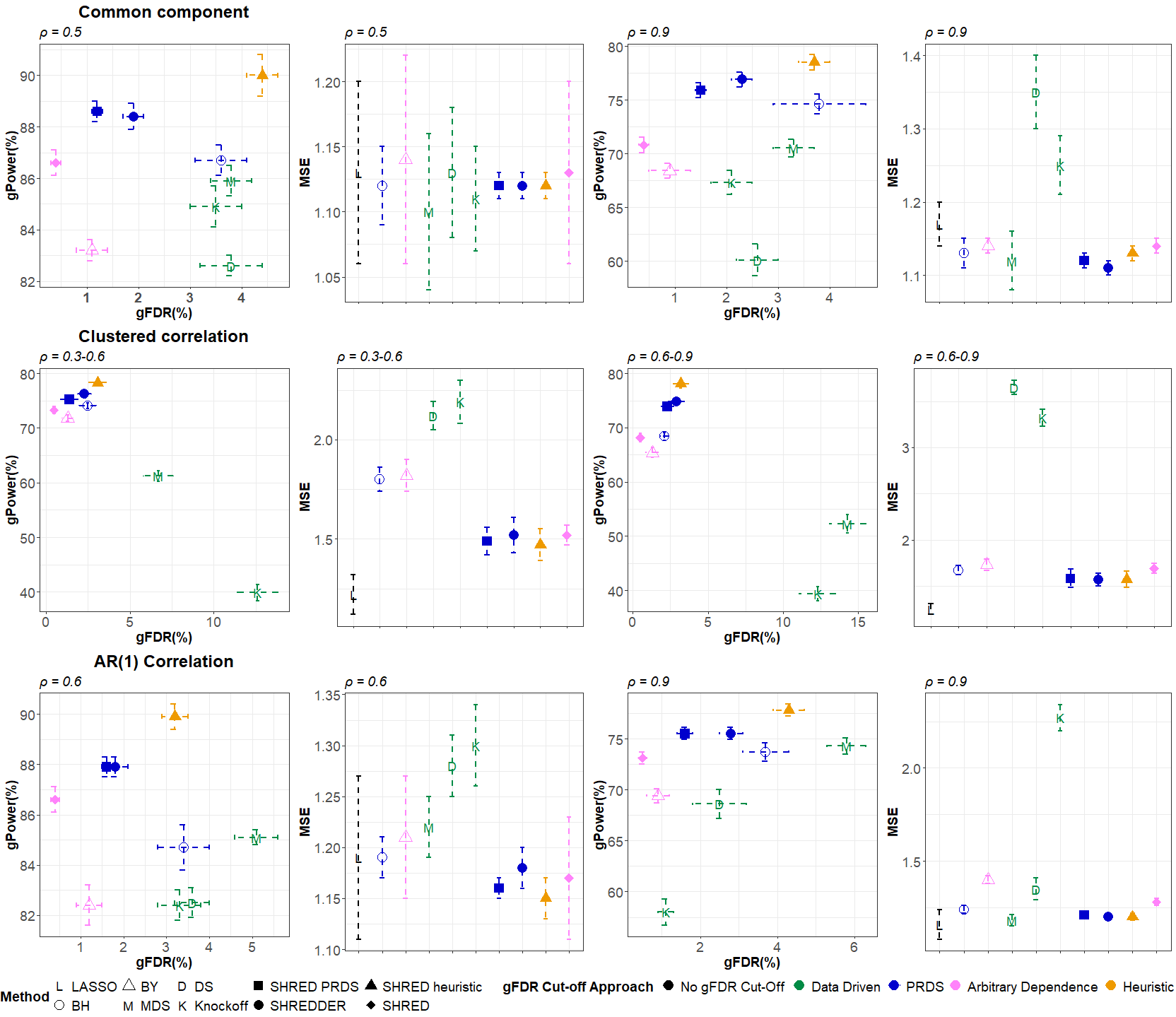}
	\caption{Gaussian regression simulation results. Error bars
          indicate one standard error.}
	\label{normalSHRED}
\end{figure}

From Figure~\ref{normalSHRED}, across all scenarios, SHRED with the
PRDS (or heuristic) cutoff and SHREDDER consistently achieve
significantly higher gPower than methods that select individual
variables, with the exception of LASSO regression, which does not
enforce FDR control. In most scenarios, SHRED with the PRDS cutoff and
SHREDDER also attain significantly lower gFDR than BH. The mirror and
knockoff methods exhibit variable performance, often showing very low
power and failing to control gFDR in the clustered-correlation
setting. SHRED outperforms BY in every scenario, achieving both higher
gPower and lower gFDR.

Although the PRDS assumptions are not guaranteed to hold in these
simulations, methods designed to control gFDR under PRDS do control
gFDR at the nominal level in practice. As expected, methods that
control gFDR without any dependence assumptions on the p-values
exhibit much tighter practical gFDR control --- often well below the
target level—at the cost of reduced power. In some scenarios, however,
the ability of SHRED to select sets of variables allows it to achieve
gPower comparable to BH while maintaining substantially lower gFDR.

Regarding predictive performance, the methods are similar in MSE under
lower or common-component correlation structures. At higher
correlations, the SHRED methods --- particularly SHRED with the PRDS
cutoff --- tend to outperform the other approaches. In nearly all cases,
the FDR-controlling procedures do not surpass the LASSO in terms of
MSE. We performed one-sample $t$-tests to determine significance of
difference in MSE for all simulations. Results are in the
Supplementary Appendix, Table~\ref{App_tab41}. In most cases the
difference in MSE is not significant, but for the clustered
correlation, at low correlation, and the AR(1) correlation with
$\rho=0.9$, SHRED-PRDS and SHREDDER both significantly outperform BH,
while SHRED significantly outperforms BY. For logistic regression,
there are also a number of cases where the SHRED methods significantly
outperform the corresponding BH or BY methods.

A detailed examination of the SHRED-PRDS and SHREDDER results
indicates that the majority of selected sets are singletons, with the
number of true singletons closely matching those identified by
BH. Notably, the number of falsely selected singletons is lower than
the number of false selections produced by BH. The remaining selected
sets are mostly of size~2, and the proportion of sets larger than
size~1 increases in scenarios with stronger predictor correlation.

We also conducted a simulation study in which the predictors were
generated independently from a standard Gaussian distribution. The
results are in Supplementary Appendix~\ref{App_SupTables}. Because
traditional FDR-controlled variable-selection methods typically
struggle when predictors are correlated, we would expect BH and BY to
perform well in this independent-predictor setting, while the larger
$\alpha$ values required by SHRED to accommodate tests of sets of
predictors might reduce its power. However, even when predictors are
simulated independently, we observe no significant difference between
SHRED and BY, nor between SHRED-PRDS and BH.

There are two reasons why our method does not perform noticeably worse
than BH or BY, even in the independent scenario. Firstly, the increase
in $\alpha$ due to testing sets of predictors is modest. Because the
additional sets receive smaller weights and are fewer in number than
the singletons, the increase in $\alpha$ is limited ---
$\sim 33\%$ under a perfect binary tree --- so even if no
multi-predictor sets are selected, the effective FDR control level
decreases by only $\sim 25\%$. Secondly, even when predictors are truly
independent, moderate sample correlations can still arise between some
pairs, and these can create challenges for identifying true
predictors.

The stong performance of SHRED under dependent conditions and
competitive performance under independent conditions demonstrates that
SHRED can be applied broadly to variable selection problems without
concern for the underlying correlation structure among predictors.

\subsubsection{Logistic Regression}

\begin{figure}[htbp]
	\centering
	\includegraphics[width=13cm, height = 10.5cm]{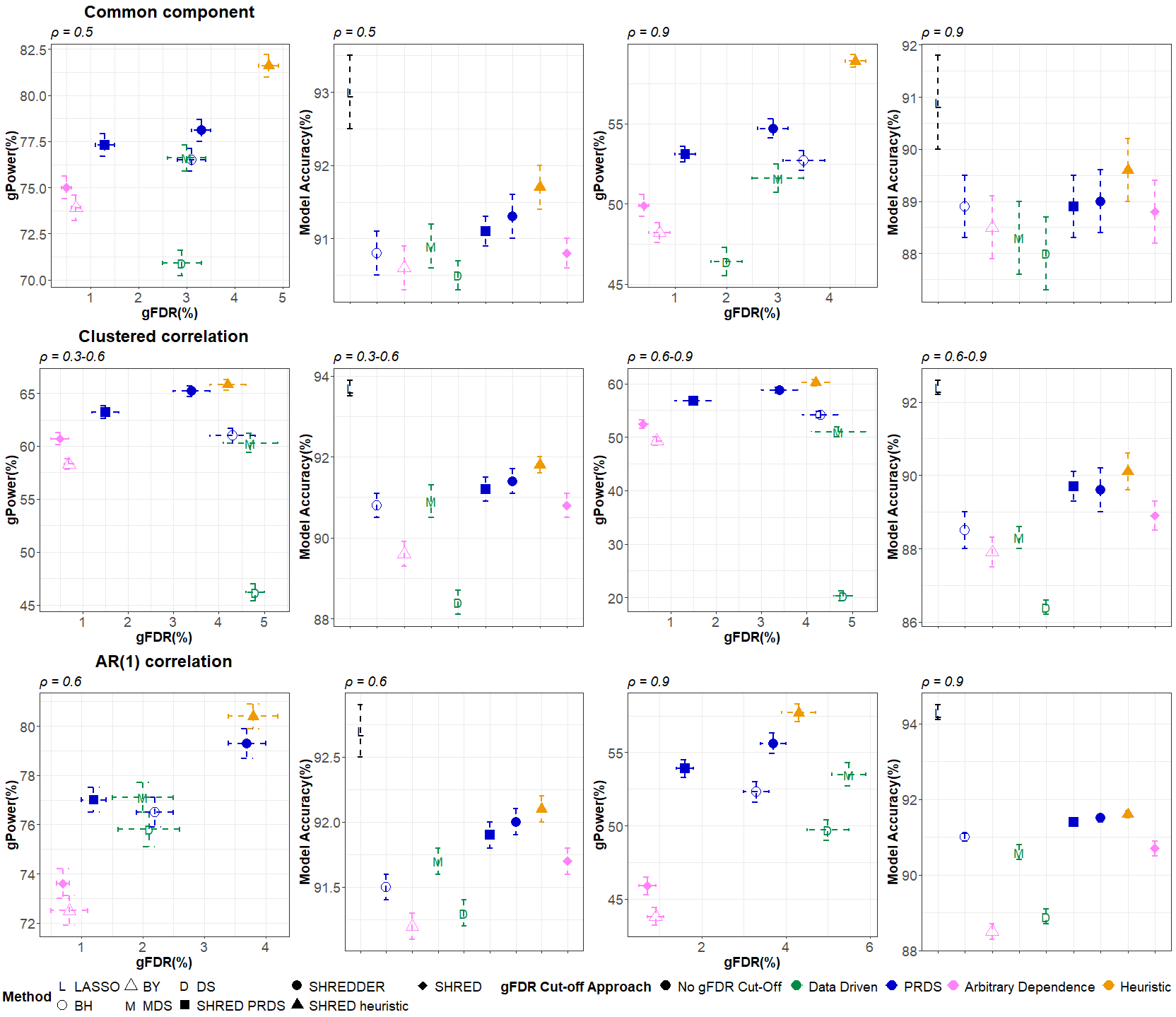}
	\caption{Logistic regression simulation results. Error bars
          indicate one standard error.}
	\label{logisticSHRED}
\end{figure}

For logistic regression (Figure~\ref{logisticSHRED}), all methods show
lower gPower than in the Gaussian regression setting, as a binomial
response contains less information than a Gaussian
response. Nevertheless, the relative performance of the methods
closely mirrors the Gaussian case: SHRED with the PRDS cutoff and
SHREDDER generally achieve higher gPower than BH, often with lower
gFDR, and their gFDR remains below the nominal level in all
scenarios. As expected, none of the FDR-controlling methods match the
power of the LASSO.

SHRED consistently shows higher gPower and lower gFDR than BY,
although the standard errors are large enough that differences are
often not significant in individual scenarios. The improved gPower and
gFDR lead to better classification accuracy. This improvement is
significant in many scenarios (see Supplementary
Table~\ref{App_tab41}). In particular, in both AR(1) scenarios,
SHRED-PRDS and SHREDDER have significantly higher accuracy than BH,
while SHRED has significantly higher accuracy than BY. In many cases,
SHRED even outperforms BH despite its lower gPower.  The mirror method
maintains FDR control in all scenarios but exhibits highly variable
power; using multiple data splits increases its power to levels
comparable to BH.

As in the Gaussian case, SHRED methods predominantly select singletons, with a small number of pairs and very few larger sets. The
numbers of true singletons selected by SHRED\_PRDS and SHREDDER do not
differ significantly from the number of true variables selected by BH.

We also conducted a logistic-regression simulation with independent
predictors and found that, as in the Gaussian case, SHRED methods did
not perform significantly worse than the corresponding BH or BY
procedures. The results are provided in Supplementary
Appendix~\ref{App_SupTables}.

\subsubsection{Poisson Regression}

\begin{figure}[htbp]
  \centering \includegraphics[width=13cm, height=10.5cm]{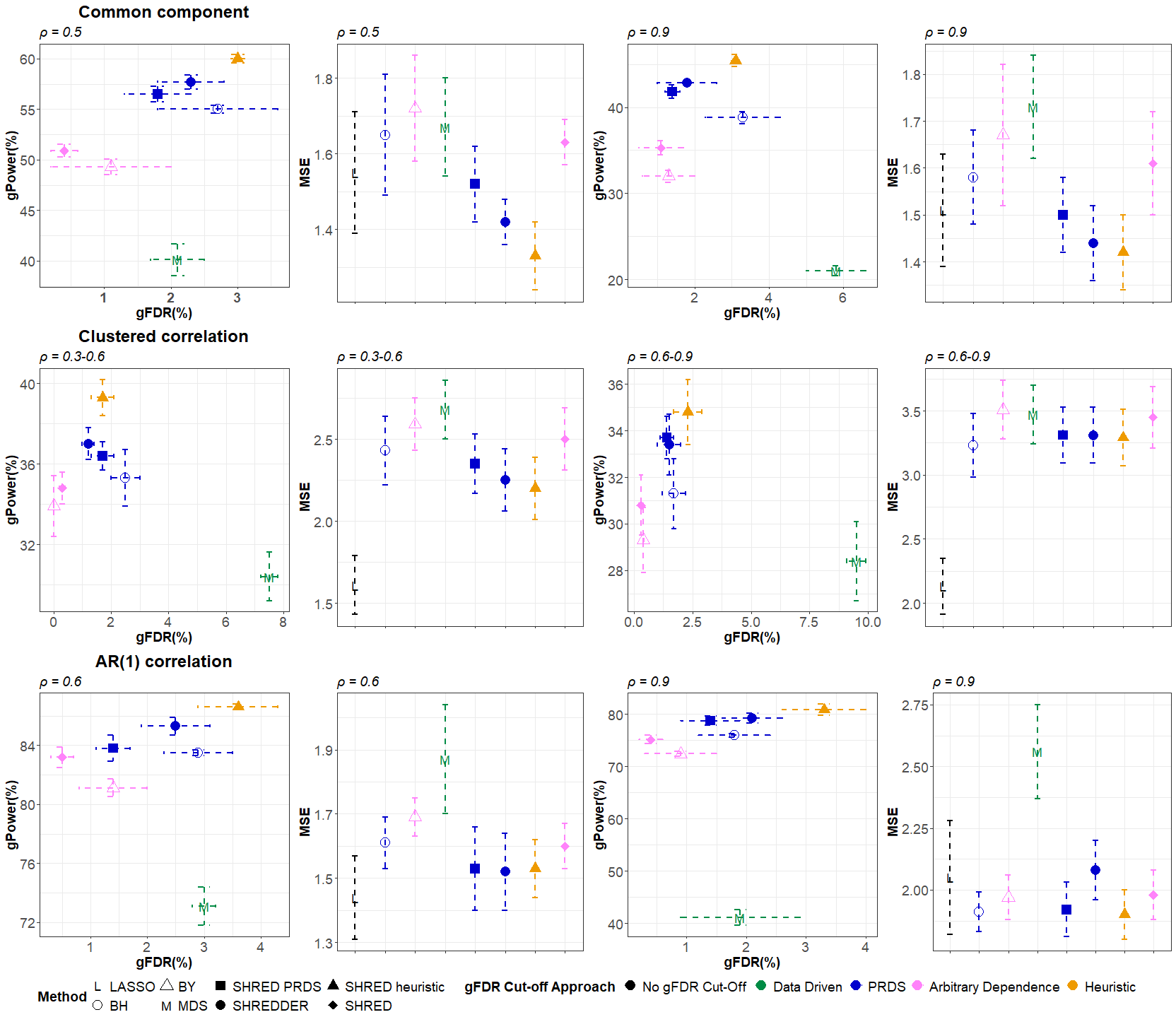}
  \caption{Poisson regression simulation results. Error bars
          indicate one standard error.}
  \label{poissonSHRED}
\end{figure}

For the Poisson regression simulations (Figure~\ref{poissonSHRED}),
all methods showed even lower gPower than for logistic regression, for
both the common-component and clustered-correlation
scenarios. Nevertheless, the relative performance
followed similar patterns to the other simulations. SHRED typically
achieved higher gPower and lower gFDR than BY, and SHRED\_PRDS and
SHREDDER generally showed higher gPower and lower gFDR than BH. The
mirror method with MDS was again unreliable, often performing poorly
and frequently failing to control FDR.

As in the Gaussian and logistic cases, no FDR-controlling
method matched the power of the LASSO. In the common-component and
clustered-correlation scenarios, the MSE of the LASSO was
significantly lower than all FDR control methods. However, in the
AR(1) correlation scenarios, the MSEs of SHRED\_PRDS
and SHREDDER were comparable to that of the LASSO. The MSEs for
SHRED\_PRDS and SHREDDER were generally smaller than for BH, but
the difference was only significant for
SHREDDER in the single common component, $\rho=0.5$ case (see Supplementary
Table~\ref{App_tab41}).

Similar to earlier results, SHRED methods primarily selected
singletons, with a few size-2 sets and very few larger
sets. The numbers of true singletons selected by SHRED PRDS and
SHREDDER did not differ significantly from the number of true
variables selected by BH.

We also conducted a Poisson-regression simulation with independent
predictors and found that, as for the other settings, SHRED
methods did not perform significantly worse than the corresponding BH
or BY procedures. The results are provided in Supplementary
Appendix~\ref{App_SupTables}.

\subsection{Simulation Based on Real Microbiome Data}

We also evaluated the performance of our methods on data exhibiting
real-world correlation structures using the MIDAS 4 global wastewater
taxonomy dataset~\citep{midas}. This dataset contains $n=1,278$
microbiome samples collected from wastewater treatment plants
worldwide, with taxon abundances for $p=604$ taxa (at genus level)
reported as whole-number read counts per sample. These counts
represent absolute sequencing measures, indicating the number of reads
assigned to each amplicon sequence variant (ASV) or taxon.

For our analysis, we first converted the raw counts to relative
abundances and then applied the centered log-ratio (CLR)
transformation, following recommendations in the microbiome literature
for addressing differences in sequencing
depth~\citep{quinn2019compositions,yerke2024proportion}. Zero counts
were handled by adding a small constant ($10^{-6}$) to each count
prior to transformation. Using the CLR-transformed relative abundances
of each genus as predictors, we simulated 100 replicates of a normally
distributed response variable $Y$. For each replicate, we randomly
selected $T=200$ true signals, then simulated $Y \sim N(X\beta,1)$,
where the $\beta$ coefficients for the true signals were generated
independently from a $N(0,1)$ distribution. Variable selection was
performed on the full dataset, and mean squared error (MSE) was
estimated via 10-fold cross-validation. For all methods, the target
gFDR level was set to $q = 0.05$.

\begin{table}
  \caption{SHRED Procedures on Microbiome $X$ data with simulated Gaussian $Y$\label{midassimshred}} 

  {
    \centering
\begin{tabular}{cccccccc}
  \hline
  \textbf{Method} & \textbf{gFDR Cut-off} & \textbf{gPower} & \textbf{SE gPower} & \textbf{gFDR} & \textbf{SE gFDR} & \textbf{MSE} & \textbf{SE MSE} \\
  \hline
  LASSO & No gFDR Cut-off &88.6\% & 0.2\% & 46.3\% & 0.9\% & 1.17 & 0.04\\
  \hline
  Mirror & MDS (Data driven) & 70.2\% & 0.5\% & 6.7\% & 0.2\% & 1.53 & 0.06\\
  Mirror & DS (Data driven) & 66.5\% & 0.5\% & 3.3\% & 0.2\% & 2.28 & 0.10\\
  Knockoff & Data driven & 69.0\% & 0.8\% & 3.7\% & 0.3\% & 2.68 & 0.05\\
  \hline
  BH & PRDS &  82.3\% & 0.7\% & 3.6\% & 0.4\% & 1.33 & 0.03 \\
  SHRED & PRDS & 85.5\% & 0.3\% & 1.5\% & 0.2\% & 1.29 & 0.02\\ 
  SHRED & Heuristic & 88.3\% & 0.2\% & 4.2\% & 0.5\% & 1.18 & 0.05\\
  SHREDDER & PPRDS & 85.8\% & 0.5\% & 3.5\% & 0.4\% & 1.22 & 0.04\\
  \hline
  SHRED & Arbitrary &80.5\% & 0.4\% & 0.2\% & 0.1\% & 2.05 & 0.08\\
  BY & Arbitrary & 77.7\% & 0.8\% & 0.7\% & 0.3\% & 1.87 & 0.08\\
  \hline
\end{tabular}
}

\end{table}

Results are shown in Table~\ref{midassimshred}. Excluding SHRED with
the heuristic cut-off, the best-performing method in terms of gPower
--- while maintaining gFDR control --- is SHREDDER. Although both
SHRED with the PRDS cut-off and SHREDDER achieved significantly higher
gPower than BH, only SHREDDER showed a significant improvement in MSE
compared with BH.

\subsection{Potential Adjustments for Further Increasing the Selection Performance of the SHRED Methods}

The idea behind SHRED is to achieve better gPower by testing sets of
surrogate variables where it is difficult to distinguish the true
variable. The trade-off is that testing additional hypotheses forces
us to increase $\alpha$, potentially reducing power. Thus, testing
sets of predictors with low correlation might degrade performance. We
may be able to improve performance by cutting the hierarchical tree at
a specified correlation value, i.e. only testing sets with a
correlation above a certain value. This allows the methods to only
test the singletons and the smaller sets that contain highly
correlated surrogate variables, thereby decreasing $\alpha$. We
simulate data using the design from Section~\ref{SecSimDes} for Gaussian
regression with an autoregressive (AR(1)) correlation structure of
magnitude $\rho = 0.6, 0.9,$ and $0.99$. For each scenario, we
simulate 100 replicates. At each correlation cut-off, the gPower is
compared to the standard BH method.

\begin{figure}[htbp]
	\centering
	\includegraphics[width=12cm, height = 5cm]{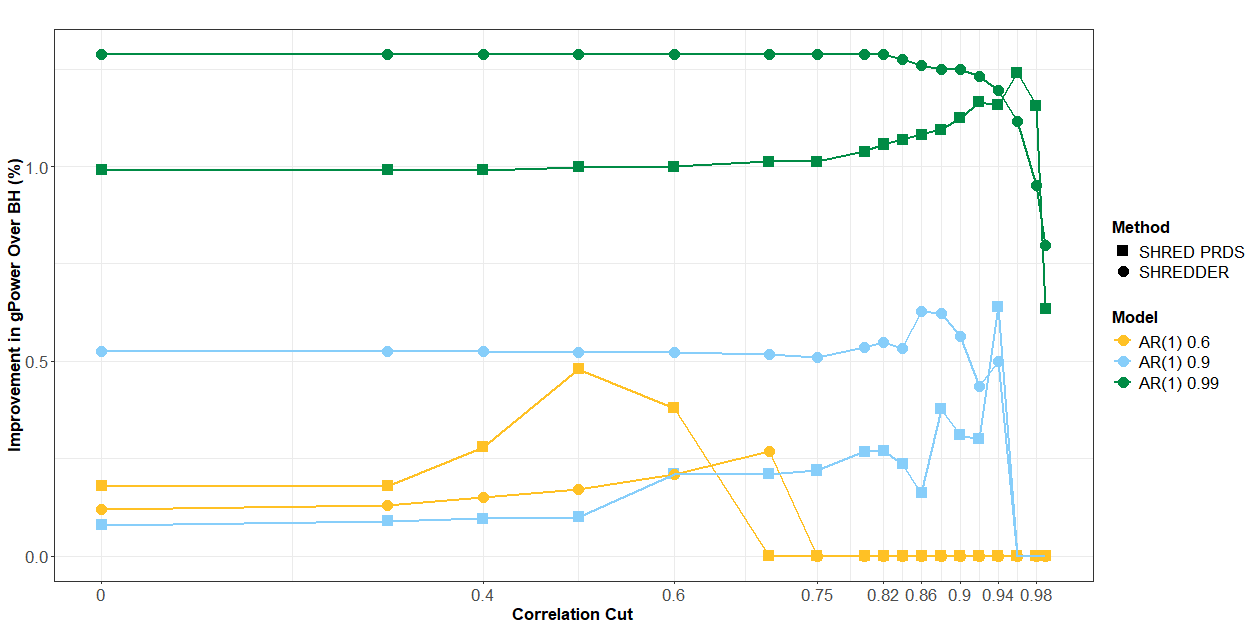}
	\caption{Difference in gPower between SHRED methods and BH, at various correlation cuts of the hierarchical tree, $p = 300, n = 1000, T = 100$.}
	\label{corcut}
\end{figure}

Figure~\ref{corcut} shows some potential to further improve gPower by
cutting the hierarchical tree. However the improvements are relatively
small, the optimal cutting point varies between scenarios, and many
wrong cutting points do worse than using the whole hierarchical tree,
so more work is needed before we can make any recommendations about
this approach.

\section{Real Data Application}\label{SecReal}

We study the MIDAS global wastewater taxonomy
dataset~\citep{midas}. As in the simulations performed using this
data, we apply the Centered Log-Ratio (CLR) transformation to the
relative abundance, following recommendations in existing
literature~\citep{quinn2019compositions,yerke2024proportion}.  Zero
counts were handled by adding a small constant ($10^{-6}$) to each
count. \citet{midas} found that the overall microbial community was
strongly affected by continent. This is assumed to be due to the
unbalanced sampling of the wastewater treatment plants and confounded
by the effects of the differing plant types and industrial
loads. Principal coordinate analysis of the Bray-Curtis beta diversity
on the samples suggested that the communities of the samples obtained
from Europe and North America were not overly distinct from one
another while these communities were distinct from those in other
continents. Therefore, we fit a logistic regression on the taxonomic
data predicting whether the continent the sample originated from was
in Europe/North America or not. To fit the logistic regression, the
genus level taxonomic data was reduced to $p = 162, n = 1278$ by
removing any rare or conditionally highly abundant taxa. For all
methods, $q$ was set to $0.05$. To assess model accuracy the data was
split into a training and test set constituting $90\%$ and $10\%$ of
the data respectively. The default accuracy associated with assigning
every sample to the largest class for the test set was $72.4\%$.

\begin{table}
  \caption{\label{midasshred} Model selection and performance for real data application}
  \centering
  \small
  \begin{tabular}{ccccc}
    \hline
    \textbf{Method} & \textbf{FDR Cut-off} & \boldmath $\sigma(I_{c_{\textrm{max}}})$ & \textbf{ Accuracy} & \textbf{LL} \\
    \hline
    LASSO & No FDR cut-off & 116 & 88.5\% & -139.1\\
    \hline
    Mirror & MDS (Data driven) &  8 & 76.5\% & -602.1\\
    Mirror & DS (Data driven) & 4 & 73.4\% & -627.3\\
    Knockoff & Data Driven & 0 & NA & NA\\
    \hline
    BH & PRDS & 45 & 83.8\% & -285.8 \\
    SHRED & PRDS & 45.75 & 84.4\% & -237.4\\ 
    SHRED & Heuristic & 52 & 85.1\% & -233.9\\
    SHREDDER & PPRDS & 46.75 & 84.8\% & -235.9\\
    \hline
    SHRED & Arbitrary & 16.9 & 80.5\% & -436.9\\
    BY & Arbitrary &  15 & 79.3\% & -458.7\\
    \hline
  \end{tabular}
\end{table}

Table~\ref{midasshred} compares the number of genera selected, 
test accuracy and log-likelihood of the resulting models for various
variable selection methods. 
SHREDDER and SHRED\_PRDS select more variables than
BH, and thus achieve better test accuracy and log-likelihood. The
mirror method with MDS selects fewer taxa compared to the BH and SHRED
methods, thus achieving lower prediction accuracy and
log-likelihood. This aligns with the results from
Table~\ref{midassimshred}. We saw in simulations that the mirror
method does not always control FDR, so there is a good chance that a
proportion of the genera selected by the mirror method are false.

\begin{figure}[htbp]
	\centering
	\caption{Midas case study selected taxa from each method}
	\label{midascasestudy}
	\includegraphics[width = 15cm, height = 8cm]{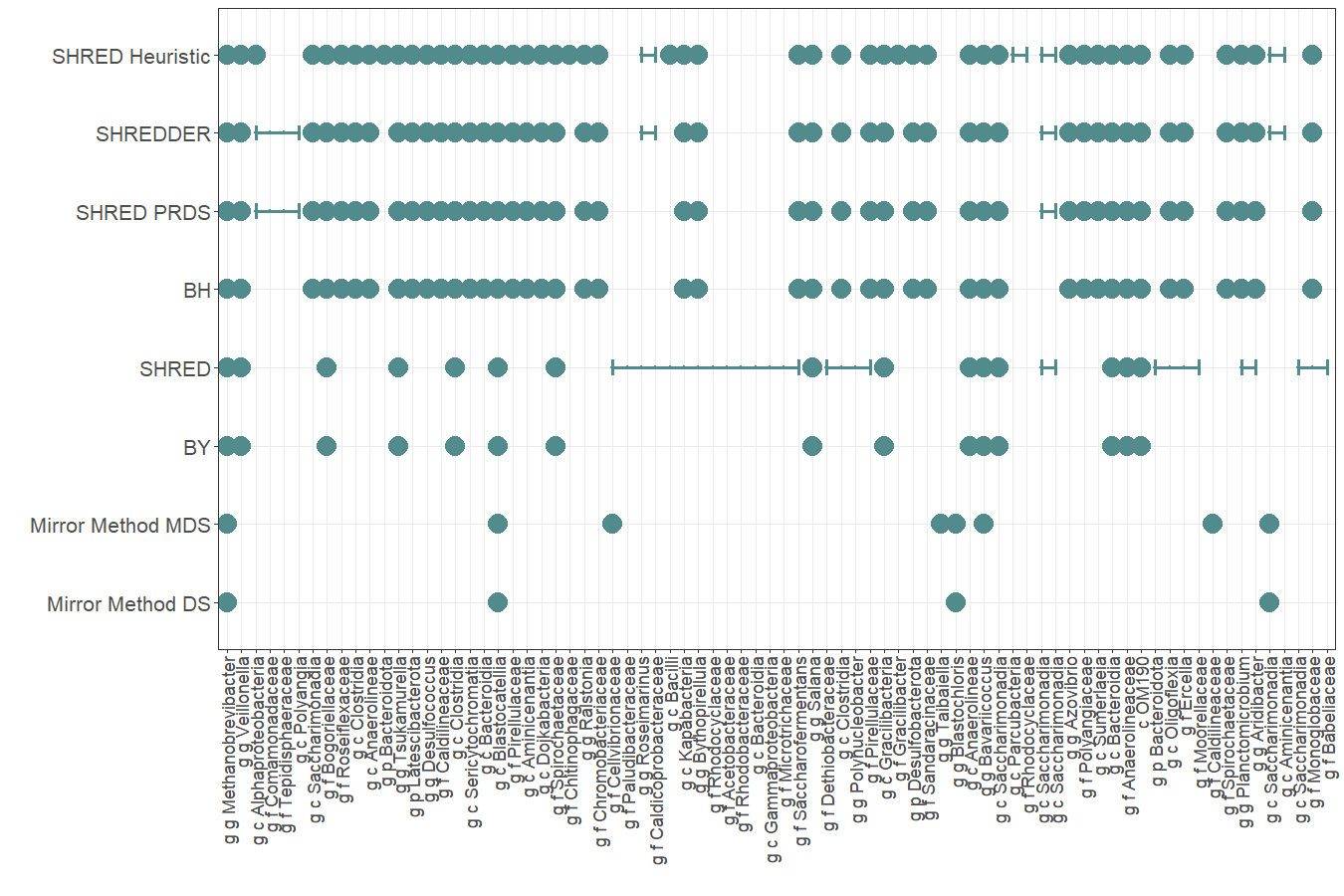}
\end{figure}

Figure~\ref{midascasestudy} shows the taxa selected by each
method. For the SHRED methods, an interval indicates a selected set of
taxa. Taxa are labelled at the lowest identifiable
taxonomic level.

The results of SHRED\_PRDS are very similar to BH, differring only in
selecting two sets of genera: four unnamed genera from different
classes and phyla; and two unnamed genera from class
Saccharimonadia. Saccharamonadia have been identified in a diverse
variety of environments, including human saliva, which suggests they
may be more abundant in domestic wastewater than industrial
wastewater. This would explain the value as a predictor: industrial
load was identified by~\cite{midas} to be a confounder for
continent. Adding one genus from each of these sets improves the model
performance on the test data.

SHREDDER selects a very similar collection of predictors, but also
selects two additional pairs of genera: one pair is the genera {\it
  Roseimarinus} and {\it Caldicoprobacteraceae}; and one pair of
unnamed genera from classes Saccharamonadia and Aminicenantia. The
mirror method identifies this genus from class Saccharamonadia as a
true predictor. The correlation between this genus and the genus from
Aminicenantia may require further investigation.

SHRED selects the same individual genera as BY, but also selects some
sets of genera. One set is the pair of genera from Saccharamonadia
selected by SHRED\_PRDS. Even using the BH cut-off, we
cannot determine which is a true predictor, but the evidence that at
least one is a true predictor is strong enough for even the strict
SHRED cut-off.  For the other sets, at least one genus is
selected by BH, supporting our claim that each set contains a true
predictor.


\section{Conclusion}\label{SecConc}
We have introduced a new framework for variable selection that allows
the selection of sets of predictors in correlated settings where it is
not possible to identify a single true predictor, but where we are
confident that at least one predictor in the set is relevant. This
addresses a key limitation of FDR-controlling methods, which often
have lower power than prediction-oriented approaches such as the
LASSO. Within this framework, we developed several variable selection
procedures based on $p$-values from hypothesis tests on sets of
variables.

The logical relationships between these hypotheses required a
substantial generalization of existing multiple testing methods,
leading to our generalized linear step-up procedure. This
generalization has broad applicability beyond the variable selection
setting. Because our SHRED and SHREDDER methods rely only
on valid conservative $p$-values, they are not restricted to
generalized linear models, unlike mirror methods and the model-X
knockoff procedure.

We proved that the generalized linear step-up procedure controls the
generalized false discovery rate (gFDR). Moreover, when the $p$-values
satisfy the PRDS condition, the rejection threshold $\alpha$ can be
reduced, allowing improved power while maintaining
gFDR control.

Extensive simulation studies demonstrate that our methods outperform
existing approaches across a range of settings. In particular, strong
performance in simulations based on real data suggests that the
practical benefits of our approach may be even greater in applied
settings.

This work introduces several conceptual advances in variable selection
and multiple hypothesis testing and opens multiple future research
directions. Firstly, while hierarchical clustering conveniently
defines sets of predictors, it may not adequately describe all
predictor structures. In non-hierarchical settings, overlapping
selected sets may arise, making the definition of an appropriate
sizing function $\sigma$ challenging. A related direction is the
development of data-dependent sizing functions, incorporating
correlation information rather than relying solely on set sizes.

Another interesting direction is adaptive versions of SHRED. Recall
that the actual gFDR control in the GLSUP is based on the sizing
function for true nulls, rather than for all hypotheses, and that we
use the value for all hypotheses as an upper bound to ensure gFDR
control. For the original BH procedure, there have been improvements
to the power by using a data-driven estimate of the number of true
nulls instead of this loose upper bound. A similar method
for the GLSUP could greatly improve power in cases with
a high proportion of false nulls.

Finally, the GLSUP framework can be extended to non-linear threshold
functions. We used the linear threshold
$\sigma\left(\overline{I_c}\right)\geqslant \alpha c$ for convenience,
but alternative threshold choices are possible. Exploring these
alternatives may lead to new classes of gFDR control methods.

\bibliography{SHRED_reference_list2.bib}
\end{document}


\maketitle

\appendix

\section{Proof of Theorems}\label{AppProofs}

In this section, we prove that the GLSUP controls the gFDR at the
nominal level. We start by establishing our notation and recalling the
GLSUP.

\subsection{GLSUP}\label{GLSUP_Method}

We let $H_1,\ldots,H_m$ be the hypotheses being tested. We let
$I_N\subseteq \{1,\ldots,m\}$ denote the (unknown) set of indices of
true null hypotheses and $I_A$ be its complement.  We let
$\mathcal{P}$ denote the powerset of $\{1,\ldots,m\}$,
i.e. $\mathcal{P} = \{J| J \subseteq \{1,\ldots,m\}\}$.

The generalized linear step-up procedure
(GLSUP) for gFDR control takes the following inputs:

\begin{itemize}
  
\item A random vector of conservative $p$-values
  $\mathbf{P}=\{P_1,\ldots,P_m\}$. (By conservative $p$-values, we mean
  that for every $i\in I_N$, the marginal distribution of $P_i$ is
  superuniform, i.e. the marginal distribution of $P_i$ stochastically
  dominates a uniform distribution, or in formulae, $(\forall
  c\in[0,1])P(P_i\leqslant c)\leqslant c$.)
  
\item A collection ${\mathcal R}\subseteq{\mathcal P}$ of
  ``rejectable'' sets of hypotheses. (This might be the whole
  powerset.) We require ${\mathcal R}$ to be closed under
  intersections and unions and contain the empty set and
  $\{1,\ldots,m\}$.  This means that for any $J\subseteq
  \{1,\ldots,m\}$, there is a smallest set $\overline{J}\in{\mathcal
    R}$ with $J\subseteq\overline{J}$. The set $\overline{J}$ is
  called the {\em closure} of $J$. We also require $I_A\in{\mathcal
    R}$. (The idea here is that $\mathcal R$ represents all sets of
  hypotheses which could plausibly be $I_A$, so we naturally assume
  whatever the true $I_A$ is, it should be in $\mathcal R$.)
  
\item A sizing function $\sigma:{\mathcal R} \rightarrow
  \mathbb{R}_{\geqslant 0}$, which is increasing with respect to set
  inclusion. This means that for any $J_1,J_2\in {\mathcal P} $ with
  $J_1\subseteq J_2$, we have $\sigma(J_1)\leqslant\sigma(J_2)$. (Note
  that we do not require $\sigma$ to be strictly increasing.)
  
\item The cut-off threshold $\alpha$. (This is calculated from the
  desired gFDR control level $q$, under particular assumptions using
  the theorems in this section.)
  
\end{itemize}

From these inputs, for any $c \in [0,1]$, we let
$I_c$ denote the set $I_c(\mathbf{P})=\{i|P_i\leqslant c\}$ and
$\overline{I}_c$ denote the set
$\overline{I}_c(\mathbf{P})=\overline{\{i|P_i\leqslant c\}}$. We
define  $c_{\textrm{max}}(\mathbf{P}) = \sup\{c \in [0,1]|
\sigma\left(\overline{I_c}\right) \geqslant \alpha c\}$. Then we reject the set
of hypotheses $\{H_i|i\in\overline{I_{c_{\textrm{max}}}}\}$.

\subsection{Proof of Theorems}\label{TheoremProofs}

We aim to prove that for the correct choice of $\alpha$, this
procedure controls gFDR at the specified level. We start with the
general case with no assumption about the dependency between $p$-values.

\begin{theorem}\label{GeneralisedBY}
  For the inputs from Section~\ref{GLSUP_Method}, suppose that for any
  $J\subseteq\{1,\ldots,m\}$, $\sigma(\overline{J})\leqslant \sum_{i\in
    J}w_i$ for some weights $w_1,\ldots,w_m$. Then the generalized
  linear step-up procedure with cut-off
  $$\alpha=\frac{\left(\sum_{i\in I_N}w_i\right)\left(1+\log(\sigma(\{1,\ldots,m\}))\right)-\sum_{i\in
      I_N}w_i\log(\sigma_i)}{q}$$ where
  $\sigma_i=\sigma(\{i\})$, controls the gFDR at level $q$.
\end{theorem}

\begin{proof}
  Let $S=I_{c_{\max}}(\mathbf{P})$ be the set of rejected hypotheses
  with small $p$-values.  Let $K=\sigma\left(\overline{S}\right)$ be the sizing
  function applied to the set of all rejected hypotheses, and let
  $V=\sigma\left(\overline{S}\cap I_N\right)$ be the sizing function of the final
  set of rejected true nulls.  As in the proof from~\cite{BY}, the false
  discovery rate is
  \begin{align*}
    {\mathbb E}(Q)&=  {\mathbb E}\left(\frac{V}{K}\right)\\
    &={\mathbb E}\left(\frac{\sigma\left(\overline{S}\cap I_N\right)}{K}\right)\\
    &\leqslant{\mathbb E}\left(\frac{\sum_{i\in S\cap I_N}w_i}{K}\right)\\
    &={\mathbb E}\left(\frac{\sum_{i\in I_N}  1_{i\in S}w_i}{K}\right)\\
    &=\sum_{i\in I_N} w_i{\mathbb E}\left(\frac{ 1_{i\in S}}{K}\right)
  \end{align*}
  where the third line follows because ${\mathcal R}$ closed under
  unions and intersections and $I_A\in{\mathcal R}$, so we
  have
  $$\overline{S}\subseteq \overline{I_A\cup (S\cap I_N)}=
  \overline{I_A}\cup\overline{(S\cap I_N)}=
  I_A\cup\overline{(S\cap I_N)}$$ Therefore
  $\overline{S}\cap I_N\subseteq \overline{(S\cap I_N)}$.  If we let
  $\phi_i$ be the measure of the marginal distribution of $P_i$, then we
  have
  \begin{align}
    \sum_{i\in I_N} w_i{\mathbb E}\left(\frac{ 1_{i\in S}}{K}\right)&=\sum_{i\in I_N} w_i\int_0^1 {\mathbb E}\left(\frac{ 1_{i\in
        S}}{K}\middle|P_i=p\right)\,d\phi_i(p)\nonumber\\
    &= \sum_{i\in I_N} w_i\int_0^1 {\mathbb E}\left(\frac{
      1_{K\geqslant\left(\alpha p\lor\sigma_i\right)}}{K}\middle|P_i=p\right)\,d\phi_i(p)\label{SecondLine}\\
    &\leqslant \sum_{i\in I_N}w_i \int_0^{\frac{\sigma(I)}{\alpha}} {\mathbb E}\left(\frac{
      1_{K\geqslant\left(\alpha p\lor\sigma_i\right)}}{K}\middle|P_i=p\right)\,d\phi_i(p)\label{ThirdLine}\\
    &\leqslant \sum_{i\in I_N} w_i\int_0^{\frac{\sigma(I)}{\alpha}} \left(\frac{1}{\alpha p}\land\frac{1}{\sigma_i}\right)\,d\phi_i(p)\nonumber\\
    &\leqslant \sum_{i\in I_N} w_i\int_0^{\frac{\sigma(I)}{\alpha}} \left(\frac{1}{\alpha p}\land\frac{1}{\sigma_i}\right)\,dp\label{FifthLine}\\
    &= \sum_{i\in I_N}
    \frac{w_i}{\alpha}\left(1+\int_{\frac{\sigma_i}{\alpha}}^{\frac{\sigma(I)}{\alpha}} \frac{1}{ p}\,dp\right)\nonumber\\
    &= \sum_{i\in I_N} \frac{w_i}{\alpha}\left(1-\log(\sigma_i)+\log(\sigma(I))\right)\nonumber\\
    &= \frac{\sum_{i\in I_N}w_i}{\alpha}\left(1+\log(\sigma(I))\right)-\frac{1}{\alpha}\sum_{i\in I_N} w_i\log(\sigma_i)\nonumber
  \end{align}
  \eqref{SecondLine} is because $c_{\textrm{max}}=\frac{K}{\alpha}$, so
  by the definition $S=I_{c_{\textrm{max}}}$, we can only have $i\in S$
  if $P_i\leqslant \frac{K}{\alpha}$. \eqref{ThirdLine} is because
  $K\leqslant \sigma(I)$, so if $P_i>\frac{\sigma(I)}{\alpha}$, then
  $i\not\in S$. \eqref{FifthLine} is because the marginal distribution of
  $P_i$ is superuniform.
\end{proof}

\begin{remark}
  In the case where there is no structure between hypotheses, so
  $\sigma(J)=|J|$ is the cardinality, we set $w_i=1=\sigma_i$, to get
  $\alpha=\frac{m_0}{q}\left(1+\log(m)\right)$. As $m_0$ is unknown,
  using $m$ as an upper bound is standard practice. Then
  Theorem~\ref{GeneralisedBY} gives
  $\alpha=\frac{m}{q}\left(1+\log(m)\right)$. This is very close, but
  slightly higher than the BY method~\citep{BY}, which uses
  $\alpha=\frac{m}{q}\sum_{i=1}^m\frac{1}{i}$. The difference is because
  in the original problem in~\citep{BY}, $K$ is constrained to be an
  integer, so where we bound $\frac{1}{K}$ by $\frac{1}{\alpha p}$, it
  could be bounded by $\frac{1}{\left\lceil {\alpha p}\right\rceil}$,
  which gives the value of $\alpha$ in the BY method.
\end{remark}

If we make stronger assumptions about the dependency between the 
$p$-values, we can prove gFDR control with a smaller $\alpha$, and
therefore achieve higher power. In particular, recall the Positive
Regression Dependancy on a Subset (PRDS) condition from~\cite{BY}:

\begin{definition}[PRDS]
  A set of $p$-values $\mathbf{P}=\{P_1,\ldots,P_m\}$ satisfies the {\em
    PRDS} condition on the subset $J\subseteq \{1,\ldots,m\}$ if for any
  $i\in J$ and any up-closed set $U\subseteq
  [0,1]^{m-1}$, $P(\mathbf{P}_{\hat{i}} \in U|
  P_i = x)$ is an increasing function in $x$, where
  $\mathbf{P}_{\hat{i}}$ is $\mathbf{P}$ with the $i$th index removed.
\end{definition}

\begin{theorem}\label{PRDS}
  For the inputs from Section~\ref{GLSUP_Method}, suppose that the
  conservative $p$-values $P_1,\ldots,P_m$ satisfy the PRDS condition
  on $I_N$, and that the sizing function $\sigma:{\mathcal
    P}\longrightarrow{}{\mathbb R}_{\geqslant 0}$ is bounded by a sum
  of weights: $(\forall
  J\subseteq\{1,\ldots,m\})\left(\sigma\left(\overline{J}\right)\leqslant \sum_{i\in
    J}w_i\right)$ for some weights $w_i\geqslant 0$. Then the generalized
  linear step-up method with cut-off
  $\alpha=\frac{\sum_{i=1}^mw_i}{q}$ controls the gFDR at level $q$.
\end{theorem}

In some cases $\sigma$ is not tightly bounded by a sum of weights,
but is much more tightly bounded by a sum of pairwise weights ---
that is, we may be able to bound $\sigma\left(\overline{J}\right)\leqslant
\sum_{i,j\in J}w_{ij}$ for some weights $(w_{ij})_{i,j=1}^m$. In these
cases, we can prove gFDR control with a tighter bound if we assume a
pairwise form of the PRDS condition:

\begin{definition}[PPRDS] A set $\{P_1,\ldots,P_m\}$ of $[0,1]$-valued
  random variables satisfies the {\em PPRDS} property on the subset
  $J\subseteq\{1,\ldots,m\}$ if it satisfies the PRDS condition on $J$
  and for any $i\ne j \in J$, and any upset $U \subseteq [0,1]^{m-2}$,
  we have
  $$P(\mathbf{P}_{\hat{i},\hat{j}} \in U|P_i = x_i, P_j = x_j)$$
  is an increasing function in $x_i$ and $x_j$, where
  $\mathbf{P}_{\hat{i},\hat{j}}$ is the vector $\mathbf{P}$ of
  $p$-values with $P_i$ and $P_j$ removed.
\end{definition}

\begin{theorem}\label{PPRDS}
  For the inputs from Section~\ref{GLSUP_Method}, suppose the
  conservative $p$-values satisfy the PPRDS condition on the set of
  true nulls and the sizing function is bounded by a sum of pairwise
  weights, i.e. $(\forall
  J\subseteq\{1,\ldots,m\})\left(\sigma\left(\overline{J}\right)\leqslant
  \sum_{i,j\in J}w_{ij}\right)$ where $w_{ij}\geqslant 0$. Then the
  generalized linear step-up method with threshold
  $\alpha=\frac{\sum_{i,j\in I_N}w_{ij}}{q}$ controls the gFDR at
  level $q$.
\end{theorem} 

In fact, Theorems~\ref{PRDS} and~\ref{PPRDS} are special cases of a
more general theorem, where we bound $\sigma\left(\overline{J}\right)\leqslant
\sum_{J'\subseteq J}w_{J'}$ where $w_{J'}\geqslant 0$ is a weight
assigned to all subsets of $\{1,\ldots,m\}$. In the particular case
where $w_J$ is only non-zero for the singleton subsets, this is
equivalent to Theorem~\ref{PRDS}; while if $w_J$ is only
non-zero for sets of size at most 2, we get
Theorem~\ref{PPRDS}. [There is a slight subtlety in this case as
  Theorem~\ref{PPRDS} does not require $w_{ij}$ to be symmetric
  (meaning $w_{ij}=w_{ji}$). However, it is not too difficult to show
  that in Theorem~\ref{PPRDS}, we can replace $w_{ij}$ by the symmetric
  weights $w'_{ij}=\frac{1}{2}\left(w_{ij}+w_{ji}\right)$. ]

We therefore only need to prove this more general theorem, which we
state in Theorem~\ref{SPRDS} below.

\begin{definition}[SPRDS] A set of random variables $\mathbf{P}=
  (P_1,\ldots,P_m)\in[0,1]^m$ satisfies the {\em Setwise Positive Regression
    Dependency on a Subset} (SPRDS) condition for a collection
  ${\mathcal S}\subseteq {\mathcal P}$ of subsets of $\{1,\ldots,m\}$
  if for any $C\in{\mathcal S}$, any $v,w\in [0,1]^{|C|}$
  with $(\forall i\in C)(v_i\leqslant w_i)$ and any upward-closed
  $U\subseteq[0,1]^{m-|C|}$, we have $P(\mathbf{P}_{\widehat{C}}\in
  U|\mathbf{P}_C=v)\leqslant P(\mathbf{P}_{\widehat{C}}\in
  U|\mathbf{P}_C=w)$, where $\mathbf{P}_{\widehat{C}}$ denotes
  $\mathbf{P}$ with indices from $C$ removed and $\mathbf{P}_{C}$
  denotes the restriction of $\mathbf{P}$ to $C$.
\end{definition}

It is not hard to see that the PRDS and the PPRDS are special cases of
the SPRDS condition, taking $\mathcal S$ to be the set of singletons
or singletons and pairs respectively.

\begin{theorem}\label{SPRDS}
  For the inputs from Section~\ref{GLSUP_Method}, suppose that the
  sizing function is bounded by $\sigma\left(\overline{J}\right)\leqslant
  \sum_{J'\subseteq J}w_{J'}$ where $w_{J'}\geqslant 0$ are weights on
  subsets of $\{1,\ldots,m\}$.  Also suppose that the conservative
  $p$-values $\mathbf{P}$ satisfy the SPRDS condition for the
  collection of subsets of $I_N$ with non-zero weight.  Then the
  generalized linear step-up
  method with threshold
  $\alpha=\frac{W}{q}$, where $W=\sum_{J\subseteq\{1,\ldots,m\}}w_J$,
  controls the gFDR at level $q$.
\end{theorem}

\begin{proof}
  For a scalar $x$, we will write $\mathbf{P}_C\leqslant x$ to mean
  $(\forall i\in C)(P_i\leqslant x)$. Let
  $S=I_{c_{\max}(\mathbf{P})}(\mathbf{P})$ be the set of rejected
  hypotheses with small $p$-values. Let $K=\sigma\left(\overline{S}\right)$ be the
  sizing function of the set of all rejected nulls, and let
  $V=\sigma\left(\overline{S}\cap I_N\right)$ be the sizing function of the final
  set of rejected true nulls.  As in the proof from~\cite{BY}, for any
  $C\subseteq I_N$ with $w_C> 0$, any $x\in{\mathbb R}^+$, and
  any $\mathbf{v}:C\longrightarrow{}[0,1]$, the SPRDS gives that
  $P(K\leqslant x|\mathbf{P}_C=\mathbf{v})$ is an increasing function in
  $x$ and $\mathbf{v}$.  The false discovery rate is
  
  \begin{align*}
    {\mathbb E}(Q)&= {\mathbb E}\left(\frac{V}{K}\right)\\
    &={\mathbb E}\left(\frac{\sigma\left(\overline{S}\cap I_N\right)}{K}\right)\\
    &\leqslant{\mathbb E}\left(\frac{\sum_{J\subseteq S\cap I_N}w_J}{K}\right)\\
    &={\mathbb E}\left(\frac{\sum_{J\subseteq I_N} w_J1_{J\subseteq S}}{K}\right)\\
    &=\sum_{J\subseteq I_N} w_J{\mathbb E}\left(\frac{ 1_{J\subseteq S}}{K}\right)
  \end{align*}
  Now by definition of $c_{\max}(\mathbf{P})$ and $S$, we have that
  $\sigma\left(\overline{S}\right)=\alpha c_{\max}(\mathbf{P})$.  We have $i\in S$ if and only
  if $P_i\leqslant c_{\max}(\mathbf{P})=\frac{K}{\alpha}$.  Therefore
  $$ {\mathbb E}(Q)\leqslant\sum_{J\subseteq I_N} w_J{\mathbb E}\left(\frac{ 1_{J\subseteq
      S}}{K}\right)=\sum_{J\subseteq I_N} w_J {\mathbb E}\left(\frac{
    1_{\left(\mathbf{P}_J\leqslant\frac{K}{\alpha}\right)}}{K}\right)$$
  Since $K=\sigma\left(\overline{S}\right)$, where $S\subseteq \{1,\ldots,m\}$,
  there are only finitely many possible values for $K$. Let the possible
  values be $k_0\leqslant k_1\leqslant\cdots\leqslant k_M$. Thus,
  \begin{adjustwidth}{-4cm}{-4cm}
    \begin{align*}
      {\mathbb E}\left(\frac{ 1_{\mathbf{P}_J\leqslant
	  \frac{K}{\alpha}}}{K}\right)
      &=\sum_{l=0}^M \frac{P\left(K=k_l\cap \left(\mathbf{P}_J\leqslant \frac{k_{l}}{\alpha}\right)\right)}{k_{l}}\\
      &=\sum_{l=0}^M \frac{P\left(\mathbf{P}_J\leqslant \frac{k_{l}}{\alpha}\right)}{k_{l}}\left(P\left(K\leqslant
      k_{l}\middle | \mathbf{P}_J\leqslant \frac{k_{l}}{\alpha}\right)-P\left(K\leqslant
      k_{l-1}\middle | \mathbf{P}_J\leqslant \frac{k_{l}}{\alpha}\right)\right)\\
    \end{align*}
  \end{adjustwidth}

For any $l$, the set of $p$-values $\{\mathbf{P}|K(\mathbf{P})\leqslant k_{l}\}$
is an upward-closed set, so by the SPRDS property, 
$$P\left(K\leqslant
k_{l-1}\middle | \mathbf{P}_J\leqslant \frac{k_{l}}{\alpha} \right)\geqslant P\left(K\leqslant
k_{l-1}\middle | \mathbf{P}_J\leqslant \frac{k_{l-1}}{\alpha}\right)$$
and therefore, 
\begin{adjustwidth}{-4cm}{-4cm}
  \begin{align*}
{\mathbb E}\left(\frac{ 1_{\mathbf{P}_J\leqslant
      \frac{K}{\alpha}}}{K}\right)&\leqslant 
\sum_{l=0}^M \frac{P\left(\mathbf{P}_J\leqslant \frac{k_{l}}{\alpha}\right)}{k_{l}}\left(P\left(K\leqslant
k_{l}\middle | \mathbf{P}_J\leqslant \frac{k_{l}}{\alpha}\right)-P\left(K\leqslant
k_{l-1}\middle | \mathbf{P}_J\leqslant \frac{k_{l-1}}{\alpha}\right)\right)\\
  \end{align*}
\end{adjustwidth}
Since $\alpha=\frac{W}{q}$ and $P_i$ are all superuniform, we have $
\frac{P\left(\mathbf{P}_J\leqslant \frac{k_{l}}{\alpha}\right)}{k_{l}}\leqslant\frac{\left(\frac{k_l}{\alpha}\right)}{k_l}=\frac{1}{\alpha}=\frac{q}{W}$.
Letting $\pi^J_l=P\left(K\leqslant k_{l}\middle |
\mathbf{P}_J\leqslant \frac{k_{l}}{\alpha}\right)$ and $\pi_{-1}^l=0$,
we have
$${\mathbb E}\left( \frac{1_{\mathbf{P}_J\leqslant \frac{K}{\alpha}}}{K}\right)\leqslant   \sum_{l=0}^M  \frac{q}{W}\left(\pi_{l}^J-\pi_{l-1}^J\right)=\frac{q}{W}\pi_{M}^J=\frac{q}{W}
$$
Thus,
$$ {\mathbb E}(Q)\leqslant\sum_{J\subseteq I_N} w_J {\mathbb E}\left(\frac{
    1_{\left(\mathbf{P}_J\leqslant\frac{K}{\alpha}\right)}}{K}\right)\leqslant\sum_{J\subseteq I_N} w_J \frac{q}{W}=q$$

\end{proof}

It is easy to see how letting $w_J$ be zero except for singletons in
Theorem~\ref{SPRDS} gives Theorem~\ref{PRDS}. We briefly elaborate on
the details of proving Theorem~\ref{PPRDS} from Theorem~\ref{SPRDS}.

\begin{proof}[Proof of Theorem~\ref{PPRDS}]
  For $J'\subseteq\{1,\ldots,m\}$, let $$W_{J'}=\left\{
  \begin{array}{ll}
    w_{ii}&\textrm{if }J'=\{i\}\\
   w_{ij}+w_{ji}&\textrm{if }J'=\{i,j\}\\
    0&\textrm{otherwise}
  \end{array}\right.$$
  We see that for any $J\subseteq\{1,\ldots,m\}$,
$$\sigma\left(\overline{J}\right)\leqslant \sum_{i,j\in J}w_{ij}=\sum_{i\in
    J}w_{ii}+\sum_{i<j\in J} (w_{ij}+w_{ji})=\sum_{i\in
    J}W_{\{i\}}+\sum_{i<j\in J} W_{\{i,j\}}=\sum_{J'\subseteq J}W_{J'}$$
Furthermore, $W_{J'}$ is zero for all subsets of size greater than 2,
so the PPRDS condition on $I_N$ implies the SPRDS condition for
subsets of $I_N$ of size at most two. Thus, by Theorem~\ref{SPRDS},
the GLSUP with $\alpha=\frac{\sum_{J\subseteq \{1,\ldots,m\}}W_J}{q}$
controls the gFDR at level $q$. We have  
$$\sum_{J\subseteq \{1,\ldots,m\}}W_J=\sum_{i=1}^m
w_{ii}+\sum_{i<j\in\{1,\ldots,m\}}(w_{ij}+w_{ji})=\sum_{i,j=1}^{m}w_{ij}$$
\end{proof}

\subsection{Proofs that the Sizing Function for SHREDDER can be
  Bounded by a Sum of Pairwise Weights on $I\times I$.} \label{IncExcIneq}

We explained that one motivation for the modified GLSUP was that it
allowed us to bound the sizing function $\sigma:\Sigma\rightarrow
{\mathbb R}_{\geqslant 0}$ by $\sigma\left(\overline{J}\right)\leqslant \sum_{i,j\in
  J}w_{ij}$ for some weights $w_{ij}\geqslant 0$. In this section we
  elaborate more on how this applies.

\subsubsection{Application of Theorem~\ref{SPRDS}}

We begin by looking at the cases under which Theorem~\ref{SPRDS} can
be applied. In order to apply Theorem~\ref{SPRDS}, we need to find
when a bound on the sizing function can be expressed as a sum of
weights on subsets. That is, if we let
$\phi(A)\geqslant\sigma\left(\overline{A}\right)$, when can we write
$\phi\left(A\right)=\sum_{A'\subset A}w_{A'}$ for weights
$w_{A'}\geqslant 0$? It turns out that this can be neatly expressed as
functions that satisfy the inclusion-exclusion inequality.

\begin{proposition}~\label{DownMeasure}
Let $\mathcal{R}$ be a collection of subsets of $\{1,\ldots,m\}$
closed under unions and intersections. An increasing set function
$\phi:\mathcal{R}\rightarrow{\mathbb R}_{\geqslant 0}$ is of the form
$\phi(A)=\sum_{A'\subseteq A}w_{A'}$ for some weights
$w_{A'}\geqslant 0$ if and only if $\phi$ satisfies the
inclusion-exclusion inequality:
  $$\phi\left(\bigcup_{n=1}^k A_n\right)\geqslant\mspace{-20mu}
\sum_{\emptyset\ne K\subseteq\{1,\ldots,k\}}
\mspace{-20mu}  (-1)^{|K|+1}\phi\left(\bigcap_{m\in K}A_m\right)$$
for all collections $A_1,\ldots,A_k\in\mathcal R$.

Furthermore, in this case, we have $w_{A}=0$ whenever the inequality
is an equality for some decomposition of $A$,
i.e. when there is a non-trivial decomposition
$A=\overline{A_1\cup\cdots\cup A_k}$, for which
  $$\phi\left(A\right)=\mspace{-30mu}
\sum_{\emptyset\ne K\subseteq\{1,\ldots,k\}}\mspace{-20mu}
  (-1)^{|K|+1}\phi\left(\bigcap_{m\in K}A_m\right)$$
This is equivalent to the condition
$$\phi(A)=-\mspace{-5mu}\sum_{\stacklimits{A'\in\mathcal{R}}{A'\subsetneq A}}
\mspace{-5mu}\mu(A,A')\phi\left(A'\right)$$
where $\mu$ is the M\"{o}bius function for the partially ordered
set $\{A''\in\mathcal{R}|A'\subseteq A''\subseteq A\}$, defined
recursively by $\mu(A,A)=1$, and
$\mu(A,B)=-\sum_{A'\in\mathcal{R},B\subsetneq A'\subseteq A}\mu(A,A')$. 
\end{proposition}

\begin{proof}
Firstly, if $\phi(A)=\sum_{A'\subseteq A} w_{A'}$ for some weights
$w_{A'}\geqslant 0$, then 
\begin{align*}
  \phi\left(\bigcup_{n=1}^k A_n\right)&=\mspace{-5mu}\sum_{\stacklimits{A'\in\mathcal{R}}{A'\subseteq \bigcup_{n=1}^k A_n}}\mspace{-5mu}w_{A'}\\
&\geqslant \mspace{-5mu}\sum_{\stacklimits{A'\in\mathcal{R}}{(\exists i\in 1,\ldots,k)(A'\subseteq  A_i)}}\mspace{-5mu}w_{A'}\\
&=\mspace{-5mu}\sum_{A'\in\mathcal{R}}\mspace{-5mu}w_{A'}\left(
  \sum_{\stacklimits{(\emptyset\ne K\subseteq\{1,\ldots,k\})}{(\forall i\in K)(A'\subseteq A_i)}}\mspace{-40mu}(-1)^{|K|+1}\right) \\
&=\mspace{-5mu}\sum_{A'\in\mathcal{R}}\mspace{-5mu}w_{A'}\left(\sum_{\emptyset\ne K\subseteq\{1,\ldots,k\}}\mspace{-30mu}(-1)^{|K|+1} 1_{A'\subseteq\bigcap_{i\in K}A_i}\right) \\
&=\mspace{-20mu}\sum_{\emptyset\ne
      K\subseteq\{1,\ldots,k\}}\mspace{-25mu}(-1)^{|K|+1}\left(\sum_{\stacklimits{A'\in\mathcal{R}}{A'\subseteq
        \bigcap_{i\in K}\!A_i}}\mspace{-5mu}w_{A'}\mspace{15mu}\right) \\
&=\mspace{-20mu}\sum_{\emptyset\ne
      K\subseteq\{1,\ldots,k\}}\mspace{-25mu}(-1)^{|K|+1}\phi\left(\bigcap_{i\in
        K}A_i\right)
\end{align*}


Conversely, suppose $\phi$ satisfies the inclusion-exclusion
inequality. We want to show that there are weights $w_{A}\geqslant 0$
for $A\in\mathcal{R}$ such that $\phi(A)=\sum_{A'\subseteq A}w_{A'}$
for all $A\in\mathcal{R}$. We recall the proof
of the standard result in partial orders that $w_A$ are given by
the M\"{o}bius inversion formula for the partially ordered set
$\mathcal{R}$.
$$w_A=\mspace{-5mu}\sum_{\stacklimits{A'\in{\mathcal R}}{A'\subseteq A}}
\mspace{-5mu}\mu(A,A')
%
\phi(A')$$
where $\mu(A,A')$ is the M\"{o}bius function on the partially ordered
set $\{A''\in\mathcal{R}|A'\subseteq A''\subseteq A\}$, defined
recursively by $\mu(A,A)=1$, and
$\mu(A,B)=-\sum_{\stacklimits{A'\in\mathcal{R}}{B\subsetneq A'\subseteq A}}\mspace{5mu}\mu(A,A')$. 
 The proof is a straightforward check that these $w_{A'}$ satisfy the
 required equation for every $A\in\mathcal{R}$, starting by proving by
 induction that for $A'\subseteq A$, $$\sum_{A'\subseteq A''\subseteq
   A}\mspace{-20mu}\mu(A'',A')=\left\{\begin{array}{ll}1&\textrm{if }A'=A\\
 0&\textrm{otherwise}\end{array}\right.$$
 This is clearly true for $A'=A$, while for $A'\subsetneq A$, we have
 \begin{align*}
   \sum_{A'\subseteq A''\subseteq A}\mspace{-20mu}\mu(A'',A')&=   \mu(A',A')+\mspace{-20mu}\sum_{A'\subsetneq
     A''\subseteq A}\mspace{-10mu}\left(\mspace{10mu}-\mspace{-20mu}\sum_{A'\subsetneq A'''\subseteq A''}
   \mspace{-20mu}\mu(A'',A''')\right)\\
   &=1-\mspace{-20mu}\sum_{A'\subsetneq A'''\subseteq A}\sum_{A'''\subseteq
     A''\subseteq A}\mspace{-20mu}\mu(A'',A''')\\
   &=1-\mspace{-20mu}\sum_{A'\subsetneq A'''\subseteq A}\mspace{-20mu}1_{A'''=A}\\
   &=0
 \end{align*}
Using this, we show that $w_A$ defined above satisfies the required condition:
 \begin{align*}
\sum_{\stacklimits{A'\in\mathcal{R}}{A'\subseteq A}}\mspace{-5mu}w_{A'}&=\sum_{\stacklimits{A'\in\mathcal{R}}{A'\subseteq A}}\mspace{5mu}\sum_{\stacklimits{A''\in\mathcal{R}}{A''\subseteq
  A'}}\mspace{-5mu}\mu(A',A'')\phi(A'')\\
&=\sum_{\stacklimits{A''\in\mathcal{R}}{A''\subseteq A}}\mspace{-5mu}\phi(A'')\mspace{-5mu}\sum_{\stacklimits{A'\in\mathcal{R}}{A''\subseteq
  A'\subseteq A}}\mspace{-5mu}\mu(A',A'')\\
&=\sum_{\stacklimits{A''\in\mathcal{R}}{A''\subseteq A}}\mspace{-5mu}\phi(A'')1_{A''=A}\\
&=\phi(A)
\end{align*}
We now just need to show that $w_A$ is non-negative for all
$A\in\mathcal{R}$. However, from the decomposition
$A\supseteq\bigcup\{A'\in\mathcal R|A'\subsetneq A\}$, we get that
$$\phi(A)\geqslant \mspace{-40mu}\sum_{\emptyset\neq\mathcal{S}\subseteq \{A'\in\mathcal
  R|A'\subsetneq A\}}\mspace{-40mu}(-1)^{|\mathcal{S}|+1}\phi\left(\bigcap
\mathcal{S}\right)=\mspace{-5mu}\sum_{\stacklimits{A'\in\mathcal{R}}{A'\subsetneq A}}\mspace{-5mu}\phi(A')\left(\sum_{\stacklimits{\emptyset\neq\mathcal{S}\subseteq \{A''\in\mathcal
  R|A''\subsetneq A\}}{\bigcap\mathcal{S}=A'}}\mspace{-60mu}(-1)^{|\mathcal{S}|+1}\right)$$
We can show by induction that 
$\sum_{\stacklimits{\mathcal{S}\subseteq \{A''\in\mathcal
  R|A''\subsetneq
  A\}}{\bigcap\mathcal{S}=A'}}(-1)^{|\mathcal{S}|}=\mu(A,A')$,
where the empty set is included in the sum, and its intersection is
defined to be $A$ in this context. This is clear when $A'=A$. Suppose this holds for all $A'\subsetneq
A''\subseteq A$. We have
\begin{align*}
  \mu(A,A')&=-\mspace{-20mu}\sum_{A'\subsetneq A''\subseteq A}\mspace{-20mu}\mu(A,A'')\\
  &=-\mspace{-20mu}\sum_{A'\subsetneq A''\subseteq A}\left(\sum_{\stacklimits{\mathcal{S}\subseteq \{A'''\in\mathcal
  R|A'''\subsetneq
  A\}}{\bigcap\mathcal{S}=A''}}\mspace{-40mu}(-1)^{|\mathcal{S}|}\right)\\
  &=\mspace{-30mu}\sum_{\stacklimits{\mathcal{S}\subseteq \{A'''\in\mathcal
  R|A'''\subsetneq
  A\}}{A'\subsetneq\bigcap\mathcal{S}}}\mspace{-40mu}(-1)^{|\mathcal{S}|+1}\\
  &=\mspace{-60mu}\sum_{\mathcal{S}\subseteq \{A'''\in\mathcal
  R|A'\subseteq A'''\subsetneq
  A\}}\mspace{-40mu}(-1)^{|\mathcal{S}|+1}
  \mspace{30mu}-\mspace{-50mu}\sum_{\stacklimits{\mathcal{S}\subseteq \{A'''\in\mathcal
  R|A'\subseteq A'''\subsetneq
  A\}}{\bigcap\mathcal{S}=A'}}\mspace{-40mu}(-1)^{|\mathcal{S}|+1}\\
  &=\mspace{-50mu}\sum_{\stacklimits{\mathcal{S}\subseteq \{A'''\in\mathcal
  R|A'\subseteq A'''\subsetneq
  A\}}{\bigcap\mathcal{S}=A'}}\mspace{-40mu}(-1)^{|\mathcal{S}|}
\end{align*}
Then the inclusion exclusion inequality for this decomposition gives
$w_A\geqslant 0$, with equality if and only if
$\phi(A)=-\sum_{\stacklimits{A'\in\mathcal{R}}{A'\subsetneq
  A}}\mu(A,A')\phi(A')$. From the above, this is equivalent to the
equality $$\phi(A)=\mspace{-50mu}\sum_{\emptyset\ne\mathcal{S}\subseteq
  \{A'\in\mathcal{R}|A'\subsetneq
  A\}}\mspace{-40mu}(-1)^{|\mathcal{S}|+1}\phi\left(\bigcap\mathcal{S}\right)$$ which
is the inclusion-exclusion equality for the decomposition
$A=\overline{\bigcup\{A'\in\mathcal{R}|A'\subsetneq A\}}$. We want to
show that if the inclusion-exclusion inequality for any decomposition
$A=\overline{A_1\cup\cdots\cup A_n}$ is an equality, then it is an
equality for this maximal decomposition. Let
$A=\overline{A_1\cup\cdots\cup A_n}$ satisfy
$$\phi(A)=\mspace{-30mu}\sum_{\emptyset\ne
  S\subseteq\{1,\ldots,n\}}\mspace{-20mu}(-1)^{|S|+1}\phi\left(\bigcap_{i\in S}A_i\right)$$
We will show that for any $A_{n+1}\subseteq A$, we have 
$$\phi(A)=\mspace{-30mu}\sum_{\emptyset\ne
  S\subseteq\{1,\ldots,n+1\}}\mspace{-30mu}(-1)^{|S|+1}\phi\left(\bigcap_{i\in S}A_i\right)$$
By the inclusion-exclusion inequality, we have
\begin{align*}
  \phi(A)&\geqslant \mspace{-30mu}\sum_{\emptyset\ne
  S\subseteq\{1,\ldots,n+1\}}\mspace{-30mu}(-1)^{|S|+1}\phi\left(\bigcap_{i\in S}A_i\right)\\
&= \mspace{-20mu}\sum_{\emptyset\ne
  S\subseteq\{1,\ldots,n\}}\mspace{-30mu}(-1)^{|S|+1}\phi\left(\bigcap_{i\in S}A_i\right)+\phi(A_{n+1})-\mspace{-30mu}\sum_{\emptyset\ne
  S\subseteq\{1,\ldots,n\}}\mspace{-20mu}(-1)^{|S|+1}\phi\left(A_{n+1}\cap\bigcap_{i\in S}A_i\right)\\
&= \phi(A)+\phi(A_{n+1})-\mspace{-30mu}\sum_{\emptyset\ne
  S\subseteq\{1,\ldots,n\}}\mspace{-30mu}(-1)^{|S|+1}\phi\left(\bigcap_{i\in
    S}\left(A_{n+1}\cap A_i\right)\right)\\
&\geqslant\phi(A)
\end{align*}
So the inclusion-exclusion inequality for this decomposition is also an
equality. By adding all subsets of $A$ to the decomposition, we have
that if there is any decomposition for which the inclusion-exclusion
inequality is an equality, then it is an equality for the maximal
decomposition as the union of all proper subsets of $A$.
\end{proof}

Recall that the mGLSUP starts from the data given in
Section~\ref{GLSUP_Method}, but with a specific choice of $\mathcal R$
so that after modifying the $p$-values, it is guaranteed that
$I_c\in{\mathcal R}$ for all $c\in[0,1]$. After the modification of
the $p$-values, the method proceeds in exactly the same way as the
GLSUP. We are interested in the particular case where the hypotheses
are arranged in a forest of hierarchical bifurcating trees, however we
can prove the result in a slightly more general context. We first
recall some basic order theory:

\begin{definition}\label{def_cover}
  For a partially ordered set $(X,\Leftarrow)$, an element $x\in X$
  {\em covers} an element $y\in X$ (denoted $y\prec x$) if
  $y\Leftarrow x$ and for any $z\in X$ with $y\Leftarrow z\Leftarrow
  x$, we have either $z=x$ or $z=y$. For example, if $(X,\Leftarrow)$
  has a tree structure, then a node covers only its children.
\end{definition}

We use $\Leftarrow$ instead of $\leqslant$, because in our motivating
example, the partial order is reverse implication between hypotheses.
We can prove our results for a partial order where each element covers
at most two other elements, and the set of rejectable hypotheses
consists of upwards-closed sets (upsets) with respect to the partial
order. We restrict to the case where $\sigma(S)=\sum_{i\in\Min(S)}
\phi_i$, for some weights $\phi_1,\ldots,\phi_m\geqslant 0$, where
$\Min(S)$ denotes the set of minimal elements in $S$.  We assume that
the $\phi_i$ are such that for any $A\subseteq B$,
$\sigma(A)\leqslant\sigma(B)$. That is, $\phi_i\geqslant\sum_{H_i\prec
  H_j}\phi_j$.

\begin{proposition}\label{Bifurcating}
  Let $\Leftarrow$ be a partial order on a set of hypotheses
  $H_1,\ldots,H_m$. Let $H_i\prec H_j$ denote the cover relation for
  this partial order. Suppose that for any $i$, there are at most two
  $j\in \{1,\ldots,m\}$ for which $H_j\prec H_i$.

  Let $\mathcal U$ be the collection of upward closed subsets of
  $\{H_1,\ldots,H_m\}$ with respect to the partial order
  $\Leftarrow$. Let $\sigma:{\mathcal U}\rightarrow {\mathbb
    R}_{\geqslant 0}$ be given by
  $\sigma(A)=\sum_{H_i\in\Min(A)}\phi_i$ for some $\phi_i\geqslant 0$,
and satisfy $\sigma(A)\leqslant\sigma(B)$, whenever $A\subseteq B$. Then
  there are weights $w_{ij}\geqslant 0$ for all $i,j=1,\ldots,m$ such
  that for any $A\in\mathcal U$, $\sigma(A)=\sum_{i,j\in A}w_{ij}$.  
\end{proposition}

\begin{proof}


  From Proposition~\ref{DownMeasure}, if we can show that
  $\sigma$ satisfies the inclusion-exclusion inequality on upsets,
  then we will have $\sigma(A)=\sum_{B\subseteq A}W_B$ for some
  $W_B\geqslant 0$. If we can then show that $W_B=0$ whenever $B$ has
  three or more minimal elements, then for any $i,j$ we let
  $\up(\{i,j\})$ be the upset generated by $i$ and $j$, and define
  $$w_{ij}+w_{ji}=\left\{\begin{array}{ll}0&\textrm{if }H_i\Leftarrow
  H_j\textrm{ or }H_j\Leftarrow H_i\\
  W_{\up(\{i,j\})}&\textrm{otherwise}\end{array}\right.$$ and
  $w_{ii}=W_{\up(\{i\})}$. It will follow that

\begin{align*}
  \sigma(A)&=\sum_{B\subseteq A}W_B\\
  &=\mspace{-5mu}\sum_{\stacklimits{B\subseteq A}{|\Min(B)|\leqslant 2}}\mspace{-5mu}W_B\\
  &=\mspace{-5mu}\sum_{\stacklimits{B\subseteq A}{(\exists
      i)B=\up(\{i\})}}\mspace{-5mu}W_B\mspace{10mu}+\mspace{5mu}\sum_{\stacklimits{B\subseteq A}{(\exists
      i,j)H_i\not\Leftarrow H_j,H_j\not\Leftarrow H_i,B=\up(\{i,j\})}}\mspace{-5mu}W_B\\
  &=\sum_{i\in A}w_{ii}+\sum_{i<j\in A} w_{ij}+w_{ji}\\
  &=\sum_{i,j\in A}w_{ij}
\end{align*}

For upsets $T_1,\ldots,T_n\in{\mathcal U}$, whose union is not one of
$T_1,\ldots,T_n$, we have
\begin{adjustwidth}{-4cm}{-4cm}
\begin{align*}
w_{T_1\cup\cdots\cup
T_n}&\leqslant \sigma(T_1\cup\cdots\cup
T_n)+\mspace{-20mu}\sum_{\emptyset\ne A\subseteq\{1,\ldots,n\}}\mspace{-20mu}(-1)^{|A|}\sigma\left(\bigcap_{i\in
  A}T_i\right)\\
&=\mspace{-40mu}\sum_{H_i\in\Min(T_1\cup\cdots\cup
T_n)}\mspace{-40mu}\phi_i\mspace{20mu}+\mspace{-10mu}\sum_{\emptyset\ne
  A\subseteq\{1,\ldots,n\}}\mspace{-30mu}(-1)^{|A|}\mspace{-10mu}\sum_{H_i\in\Min\left(\bigcap_{j\in  A}T_j\right)}\mspace{-40mu}\phi_i\\
&=\sum_{H_i\in T_1\cup\cdots\cup
  T_n} \mspace{-20mu}\phi_i \left(\sum_{A\subseteq\{1,\ldots,n\}}(-1)^{|A|} 1_{H_i\in\Min\left(\bigcap_{j\in
    A}T_j\right)}\right)
\end{align*}
\end{adjustwidth}
We need to show that this is always non-negative, and for any $U$ with
three minimal elements, there is a decomposition $U=T_1\cup\cdots\cup
T_n$ with $T_i\ne U$, for which this sum is zero.
We consider the term
$\sum_{\stacklimits{A\subseteq\{1,\ldots,n\}}{H_i\in\Min\left(\bigcap_{j\in
    A}T_j\right)}}(-1)^{|A|}$ for each $H_i$. We will show that
\begin{adjustwidth}{-2.5cm}{-2cm}
\begin{equation}
\sum_{\stacklimits{A\subseteq\{1,\ldots,n\}}{H_i\in\Min\left(\bigcap_{j\in
A}T_j\right)}}\mspace{-20mu}(-1)^{|A|}=\left\{
  \begin{array}{ll}
  1&\textrm{if }(\exists i_1,i_2,j_1,j_2)(H_{i_1}\prec
  H_i,H_{i_2}\prec H_i,H_{i_1}\in T_{j_1},H_{i_2}\in T_{j_2}\textrm{ and }(\forall j)\{H_{i_1},H_{i_2}\}\not\subseteq
  T_j)\\
  0&\textrm{otherwise}
\end{array}\right.
\label{EqBifurcating}
\end{equation}
\end{adjustwidth}
That is, the sum is zero unless $H_i$ covers two other hypotheses
which are in different upsets, in which case the sum is 1.  From this,
it is immediately clear that the bound is always non-negative,
i.e. the inclusion-exclusion inequality holds, which together with
$\sigma$ being an increasing function, means that $\sigma$ is
expressible as a sum of non-negative weights
$\{W_U|U\in\mathcal{U}\}$.

Suppose $H_i$, $H_j$ and $H_k$ are distinct minimal elements of the upset
$U\in\mathcal{U}$. Then we have
$U=(U\setminus\{H_i\})\cup(U\setminus\{H_j\})\cup(U\setminus\{H_k\})$, and
we have
\begin{align*}
  W_U&\leqslant
  \sigma(U)-\sigma(U\setminus\{H_i\})-\sigma(U\setminus\{H_j\})-\sigma(U\setminus\{H_k\})\\
  &\qquad\qquad+\sigma(U\setminus\{H_i,H_j\})+\sigma(U\setminus\{H_i,H_k\})+\sigma(U\setminus\{H_j,H_k\})-\sigma(U\setminus\{H_i,H_j,H_k\})
\end{align*}
This gives 
$$W_U\leqslant\sum_{H_l\in
  U}\mspace{-5mu}\phi_l\mspace{-5mu}\sum_{A\subseteq\{H_i,H_j,H_k\}}\mspace{-20mu}(-1)^{|A|}1_{H_l\in
  \min\left(U\setminus A\right)}$$

From~\eqref{EqBifurcating}, we have
$\sum_{A\subseteq\{H_i,H_j,H_k\}}(-1)^{|A|}1_{H_l\in \min\left(U\setminus
  A\right)}$ is 0 unless $H_l$ covers two elements in $U$ and every
one of $U\setminus\{H_i\}$, $U\setminus\{H_j\}$, $U\setminus\{H_k\}$
excludes one of these elements. This is clearly impossible.  Thus
$\sum_{A\subseteq\{H_i,H_j,H_k\}}(-1)^{|A|}1_{H_l\in \min\left(U\setminus
  A\right)}=0$ for all $l$, meaning $W_U=0$. Thus, the only non-zero
$W_U$ are ones where $U$ has at most two minimal elements.

We just need to prove~\eqref{EqBifurcating}. Clearly, the only nonzero
terms in the sum are those where $A$ is restricted to the $j$ for
which $H_i\in T_j$. Without loss of generality, let $\{j|H_i\in
T_j\}=\{1,\ldots,l\}$ for some $l\leqslant n$. We therefore want to
determine for which $A\subseteq\{1,\ldots,l\}$, $H_i$ is a minimal
element of $\bigcap_{j\in A}T_j$. Since $\bigcap_{j\in A}T_j$ is
upward-closed, $H_i$ is a minimal element if $\bigcap_{j\in A}T_j$
does not contain any $H_j$ covered by $H_i$. We know that $H_i$ covers
at most two elements. We consider three cases:

If $H_i$ covers no elements in $\bigcup_{i=1}^l T_i$, then it is
minimal in all downsets containing
it. $$\sum_{A\subseteq\{1,\ldots,n\}}\mspace{-15mu}(-1)^{|A|}
1_{H_i\in\Min\left(\bigcap_{j\in
    A}T_j\right)}=\mspace{-15mu}\sum_{A\subseteq\{1,\ldots,l\}}\mspace{-15mu}(-1)^{|A|}=0$$

If $H_i$ covers only one $H_j\in T_1\cup\cdots\cup T_l$, then
$H_i$ is minimal in all intersections which do not contain
$H_j$. Let $B=\{k\in\{1,\ldots,l\}|H_j\not\in T_k\}$. Then $H_i$ is
minimal in $\bigcap_{k\in A}T_k$ if and only if $A$ contains an
element in $B$.
Thus
$$\sum_{A\subseteq\{1,\ldots,n\}}\mspace{-15mu}(-1)^{|A|}
1_{H_i\in\Min\left(\bigcap_{j\in
    A}T_j\right)}=\mspace{-15mu}\sum_{\stacklimits{A\subseteq\{1,\ldots,l\}}{A\cap
    B\ne\emptyset}}\mspace{-15mu}(-1)^{|A|}
=\mspace{-15mu}\sum_{A\subseteq\{1,\ldots,l\}}\mspace{-15mu}(-1)^{|A|}-\mspace{-25mu}\sum_{A\subseteq
  \{1,\ldots,l\}\setminus B}\mspace{-25mu}(-1)^{|A|}=0$$ (We know $
\{1,\ldots,l\}\setminus B\ne\emptyset$ by our assumption $H_j\in
T_1\cup\cdots\cup T_l$.)

If $H_i$ covers two elements, $H_{i_1}$ and $H_{i_2}$ in $T_1\cup\cdots\cup
T_l$, then $H_i$ is minimal in an intersection if it does not contain
either $H_{i_1}$ or $H_{i_2}$.  Let
$B_1=\{k\in\{1,\ldots,l\}|H_{i_1}\not\in T_k\}$ and
$B_2=\{k\in\{1,\ldots,l\}|H_{i_2}\not\in T_k\}$ . Then $H_i$ is
minimal in $\bigcap_{k\in A}T_k$ if and only if $A$ contains an
element of $B_1$ and an element of $B_2$. Thus
\begin{adjustwidth}{-4cm}{-4cm}
\begin{align*}
\sum_{A\subseteq\{1,\ldots,n\}}\mspace{-15mu}(-1)^{|A|} 1_{H_i\in\Min\left(\bigcap_{j\in
    A}T_j\right)}&=\mspace{-15mu}\sum_{\stacklimits{A\subseteq\{1,\ldots,l\}}{A\cap
    B_1\ne\emptyset,A\cap B_2\ne\emptyset}}\mspace{-15mu}(-1)^{|A|}\\
&=\mspace{-15mu}\sum_{A\subseteq\{1,\ldots,l\}}\mspace{-15mu}(-1)^{|A|}-\mspace{-35mu}\sum_{A\subseteq\{1,\ldots,l\}\setminus
  B_1}\mspace{-25mu}(-1)^{|A|}-\mspace{-35mu}\sum_{A\subseteq\{1,\ldots,l\}\setminus
  B_2}\mspace{-25mu}(-1)^{|A|}\mspace{10mu}+\mspace{-35mu}\sum_{A\subseteq\{1,\ldots,l\}\setminus (B_1\cup
  B_2)}\mspace{-35mu}(-1)^{|A|}\\
&=\left\{\begin{array}{ll}1&\textrm{if }B_1\cup B_2=\{1,\ldots,l\}\\
0&\textrm{otherwise}\end{array}\right.
\end{align*}
\end{adjustwidth}
The event $B_1\cup B_2=\{1,\ldots,l\}$ is equivalent to the event that
no $T_j$ contains both $H_{i_1}$ and $H_{i_2}$.

\end{proof}

\begin{remark}
The above proof might at first glance seem to indicate that $w_U=0$
whenever $U$ is an upset with only one minimal element. However, if
$U$ has only one minimal element, it cannot be expressed as a
non-trivial union of upsets, so instead of the inclusion-exclusion
inequality, we use the fact that $\sigma$ is an increasing function,
and we have $w_U=\sigma(U)-\sigma\left(U\setminus\{H_i\}\right)$,
where $H_i$ is the unique minimal element of $U$.
\end{remark}

\subsection{Cut-offs for SHRED and SHREDDER}~\label{Shredcutoff}

In this section, we apply the theorems from
Section~\ref{TheoremProofs} to the FDR control problems arising from
the SHRED and SHREDDER methods to provide specific values for the
cut-offs there.

\begin{corollary}
  Let null hypotheses $H_1,\ldots,H_m$ be partially ordered by reverse
  implication, and let the number of discoveries resulting from the
  rejection of a subset $J\subseteq\{1,\ldots,m\}$ of the hypotheses
  be given by a sizing function $\sigma(J)\leqslant
  \sum_{i\in\Min(J)}\phi_i$ --- where $\phi_i\geqslant 0$ are weights
  on hypotheses and $\Min(J)$ denotes the set of minimal hypotheses in
  $J$ under reverse implication (i.e. the null hypotheses that do not
  imply any other hypotheses in $J$).  Let
  $I_N\subseteq\{1,\ldots,m\}$ be the set of indices of true null
  hypotheses.  Let $p=\sum_{i\in\Min(\{1,\ldots,m\})}\phi_i$ and
  $r_0=\sum_{i\in I_N}\phi_i$.  For any $q>0$, the generalized linear
  step-up procedure with
  $\alpha_{\textrm{SHRED}}=\frac{r_0(1+\log(p))-\sum_{i\in
      I_N}\phi_i\log(\phi_i)}{q}$
  controls the gFDR at level $q$.
\end{corollary}
	
\begin{proof}
  By Theorem~\ref{GeneralisedBY}, the gFDR is controlled at level $q$
  when we use the cut-off
    $$\alpha_{\textrm{SHRED}}=\frac{\left(\sum_{i\in I_N}\phi_i\right)\left(1+\log(\sigma(\{1,\ldots,m\}))\right)-\sum_{i\in
    I_N}\phi_i\log(\sigma_i)}{q}=\frac{r_0(1+\log(p))-\sum_{i\in I_N}\phi_i\log(\phi_i)}{q}$$
\end{proof}

\begin{corollary}\label{prds}
  Let null hypotheses $H_1,\ldots,H_m$ be partially ordered by reverse
  implication, and let the number of discoveries resulting from the
  rejection of a subset $J\subseteq\{1,\ldots,m\}$ of the hypotheses
  be given by a sizing function
  $\sigma(J)\leqslant\sum_{i\in\Min(J)}\phi_i$ --- where
  $\phi_i\geqslant 0$ are weights on hypotheses and $\Min(J)$ denotes
  the set of minimal hypotheses in $J$ under reverse implication.  Let
  $I_N\subseteq\{1,\ldots,m\}$ be the set of indices of true null
  hypotheses. Suppose that the conservative $p$-values satisfy the
  PRDS condition on the set of true nulls. For any $q>0$, the
  generalized linear step-up procedure with
  $\alpha_{\textrm{SHRED}_{\textrm{PRDS}}} = \frac{\sum_{i=1}^m
    \phi_i}{q}$
  controls the gFDR at level $q$.
\end{corollary}

\begin{proof}
  This immediately follows from Theorem~\ref{PRDS}.
\end{proof}

\begin{corollary}\label{shredder}
  Let null hypotheses $H_1,\ldots,H_m$ be partially ordered by reverse
  implication, in a bifurcating structure (meaning each hypothesis
  covers at most two others) and let the number of discoveries
  resulting from the rejection of a subset $J\subseteq\{1,\ldots,m\}$
  of the hypotheses be given by a sizing function
  $\sigma(J)=\sum_{i\in\Min(J)}\phi_i$ --- where $\phi_i\geqslant 0$
  are weights on hypotheses and $\Min(J)$ denotes the set of minimal
  hypotheses in $J$ under reverse implication.  Let
  $I_N\subseteq\{1,\ldots,m\}$ be the set of indices of true null
  hypotheses. Suppose the modified $p$ values $\{P'_i|i=1,\ldots,m\}$
  satisfy the PPRDS condition on the set of indices of true nulls. For
  any $q>0$, the generalized linear step-up procedure with
 $\alpha_{\textrm{SHREDDER}} =
 \frac{\sigma({1,\ldots,m})}{q}$ controls the gFDR at level $q$.
\end{corollary}

\begin{proof}
  This immediately follows from Theorem~\ref{PPRDS} and Proposition~\ref{Bifurcating}. 
\end{proof}

\section{Empirical gFDR Control of SHRED}\label{FDR_tables}

The proofs of FDR control for step-up methods consider the worst
possible case. In practice, FDR control is usually much
tighter. Furthermore, the theoretical cut-off values depend on the
unknown true nulls, whereas in practice, to choose the cut-offs, we
use all null hypotheses to obtain an upper bound. We conduct an
empirical study to determine the actual FDR for a typical
situation. To this end, we generated 100 simulated datasets with
dimensionality fixed at $p = 300$ and sample size $n = 1000$. Each
dataset was drawn from $X \sim N(\mu, \Sigma)$, where $\mu =
(0,\ldots,0)^{\mathsf T}$, $\Sigma_{ij} =\left\{\begin{array}{ll}
1&\textrm{if }i=j\\0.5&\textrm{otherwise}\end{array}\right.$. The
Gaussian response was constructed using $T = 100$ true predictors
selected at random, with corresponding regression coefficients
$\beta_i$ independently sampled from a standard normal distribution
for $i = 1,\ldots,T$. Under this simulation design, we assessed the
empirical FDR for SHRED with a variety of values of threshold slope
$\alpha$.

\begin{sidewaystable}
  \caption{\label{Table:2} SHRED
    empirical gFDR} \centering
  \begin{tabular}{rllllllcccccc}
    \hline
    \multicolumn{1}{c}{$\alpha$} &    \multicolumn{3}{c}{Nominal FDR (All Nulls)} &
    \multicolumn{3}{c}{Nominal FDR (True Nulls)} & gFDR & SE gFDR  & gPower & SE gPower & MSE & SE MSE \\
    & heur. & PRDS & SHRED & heur. & PRDS & SHRED \\
    \hline
    6,000  &  0.05  & 0.065 & 0.457 & 0.033 & 0.038 & 0.430 & 3.0\% & 0.3\% & 93\% & 0.5\% & 1.14 & 0.08\\
    6,667  &  0.045 & 0.059 & 0.411 & 0.030 & 0.035 & 0.387 & 3.0\% & 0.3\% & 92\% & 0.5\% & 1.14 & 0.08\\
    7,500  &  0.04  & 0.052 & 0.365 & 0.027 & 0.031 & 0.344 & 2.7\% & 0.3\% & 92\% & 0.5\% & 1.14 & 0.09\\
    8,571  &  0.035 & 0.046 & 0.319 & 0.023 & 0.027 & 0.301 & 2.3\% & 0.2\% & 91\% & 0.5\% & 1.15 & 0.08\\
    10,000  &  0.03 & 0.039 & 0.274 & 0.020 & 0.023 & 0.258 & 2.1\% & 0.2\% & 90\% & 0.4\% & 1.15 & 0.08\\
    12,000  &  0.025& 0.032 & 0.228 & 0.017 & 0.019 & 0.215 & 1.7\% & 0.2\% & 89\% & 0.5\% & 1.15 & 0.09\\
    15,000  &  0.02 & 0.026 & 0.182 & 0.013 & 0.015 & 0.172 & 1.4\% & 0.1\% & 88\% & 0.4\% & 1.16 & 0.09\\
    20,000  &  0.015& 0.019 & 0.137 & 0.010 & 0.012 & 0.129 & 1.0\% & 0.1\% & 88\% & 0.4\% & 1.16 & 0.09\\
    30,000  &  0.01 & 0.013 & 0.091 & 0.007 & 0.007 & 0.086 & 0.7\% & 0.1\% & 87\% & 0.4\% & 1.17 & 0.08\\
    \hline
  \end{tabular}
  
  \vspace{\baselineskip}
  \raggedright
  For SHRED and PRDS cut-offs, the exact theoretical FDR control level
  depends on the clustering pattern. Values in the table are an
  average over 100 simulations.
\end{sidewaystable}     

From Table~\ref{Table:2}, we see that for this simulation, the
empirical FDR closely matches the nominal heuristic FDR based on the
number of true nulls, and is close to, but slightly less than the PRDS
FDR. This suggests that the PRDS assumption may hold, or approximately
hold for these simulations, and that the empirical FDR is slightly
less than the nominal FDR based on true nulls. As we do not know the
number of true nulls, and use the total number of nulls as a bound,
this results in empirical FDR substantially below the nominal level.
The reason that the heuristic and PRDS cut-offs based on the true null
are close for this simulation, is that with 100 randomly chosen true
variables, a large majority of sets of size more than 1 contain at
least one true variable, and are therefore not true nulls. With almost
all true nulls at the lowest level of the hierarchical tree, we have
$\sum_{i\in I_N}\phi_i\approx p_0$.

\section{Simulation Tables}\label{SupTables}

Additional details for the simulation plots.
\subsection{Gaussian Regression}

\begin{table}[H]
  \caption{\label{tabx} Gaussian regression, $n = 1000, p = 300, T = 100$, independent predictors}
  \centering
  \small
  \begin{tabular}{llccrccc}
    \hline
    \textbf{Method} & \textbf{FDR Cut-off} & \textbf{Power} & \textbf{SE Power} & \textbf{FDR} & \textbf{SE FDR} & \textbf{MSE} & \textbf{SE MSE} \\
    \hline
    Lasso & No FDR cut-off & 97.7\% & 0.2\% & 55.6\% & 0.5\% & 1.1016 & 0.0723 \\
    \hline
    Mirror & MDS Data driven & 90.6\% & 0.9\% & 3.2\% & 0.2\% & 1.2097 & 0.0922 \\
    Mirror & DS Data driven & 87.2\% & 0.9\% & 2.3\% & 0.3\% & 1.3336 & 0.0713 \\
    Knockoff & Data driven & 92.0\% & 0.3\% & 3.8\% & 0.3\% & 1.2155 & 0.0694 \\
    \hline
    BH & PRDS & 92.9\% & 0.3\% & 3.4\% & 0.2\% & 1.1159 & 0.0634 \\
    SHRED & PRDS & 92.7\% & 0.3\% & 3.3\% & 0.2\% & 1.1261 & 0.0621 \\
    SHRED heuristic & Heuristic & 93.2\% & 0.4\% & 4.1\% & 0.3\% & 1.1201 & 0.0594 \\
    SHREDDER & PPRDS & 92.8\% & 0.3\% & 3.3\% & 0.2\% & 1.1260 & 0.0687 \\
    \hline
    SHRED & Arbitrary & 90.8\% & 0.3\% & 0.4\% & 0.1\% & 1.1545 & 0.0701 \\
    BY & Arbitrary & 90.7\% & 0.3\% & 0.6\% & 0.1\% & 1.1595 & 0.0713 \\
    \hline
  \end{tabular}
\end{table}

\begin{table}[H]
  \caption{\label{tab05} Gaussian regression, $n = 1000, p = 300, T = 100$, single common component correlation $\rho = 0.5$}
  \centering
  \small
  \begin{tabular}{llccrccc}
    \hline
    \textbf{Method} & \textbf{FDR Cut-off} & \textbf{Power} & \textbf{SE Power} & \textbf{FDR} & \textbf{SE FDR} & \textbf{MSE} & \textbf{SE MSE} \\
    \hline
    Lasso & No FDR cut-off & 96.3\% & 0.4\% & 53.1\% & 0.3\% & 1.1822 & 0.0311\\
    \hline
    Mirror & MDS (Data driven) & 85.9\% & 0.6\% & 3.8\% & 0.4\% & 1.1009 & 0.0623\\
    Mirror & DS (Data driven) & 81.6\% & 0.4\% & 3.8\% & 0.6\% & 1.1343 & 0.0496\\
    Knockoff & Data driven & 84.9\% & 0.8\% & 3.5\% & 0.5\% & 1.1138 & 0.0430\\
    \hline
    BH & PRDS & 86.7\% & 0.6\% & 3.6\% & 0.5\% & 1.1251 & 0.0297 \\
    SHRED & PRDS & 88.6\% & 0.4\% & 1.2\% & 0.1\% & 1.1273 & 0.0187\\ 
    SHRED & Heuristic & 90.0\% & 0.5\% & 4.4\% & 0.3\% & 1.1202 & 0.0118\\
    SHREDDER &PPRDS & 88.4\% & 0.5\% & 1.9\% & 0.2\% & 1.1214 & 0.0114\\
    \hline
    SHRED & Arbitrary & 85.5\% & 0.5\% & 0.1\% & 0.09\%\hspace{-0.6em} &1.1337  & 0.0079\\
    BY & Arbitrary & 83.2\% & 0.4\% & 1.1\% & 0.3\% & 1.1396 & 0.0897\\
    \hline
  \end{tabular}
\end{table}

\begin{table}[H]
  \caption{\label{tab07} Gaussian regression, $n = 1000, p = 300, T = 100$, single common component correlation $\rho = 0.9$}
  \centering
  \small
  \begin{tabular}{llccrccc}
    \hline
    \textbf{Method} & \textbf{FDR Cut-off} & \textbf{Power} & \textbf{SE Power} & \textbf{FDR} & \textbf{SE FDR} & \textbf{MSE} & \textbf{SE MSE} \\
    \hline
    Lasso & No FDR cut-off & 94.8\% & 0.06\% & 50.7\% & 2.8\% & 1.1701 & 0.0302\\
    \hline
    Mirror & MDS (Data driven) &70.5\% & 0.8\% & 3.3\% & 0.4\% & 1.1225 & 0.0413\\
    Mirror & DS (Data driven) &60.1\% & 1.5\% & 2.6\% & 0.4\% & 1.3498 & 0.0524\\
    Knockoff & Data driven &  67.3\% & 1.1\% & 2.1\% & 0.4\% & 1.2525 & 0.0407\\
    \hline
    BH & PRDS & 74.6\% & 0.9\% & 3.8\% & 0.9\% & 1.1313 & 0.0093 \\
    SHRED & PRDS & 75.9\% & 0.7\% & 1.5\% & 0.1\% & 1.1276 & 0.0119\\ 
    SHRED & Heuristic &78.5\% & 0.7\% & 3.7\% & 0.3\% & 1.1101 & 0.0095\\
    SHREDDER & PPRDS & 76.9\% & 0.7\% & 2.3\% & 0.2\% & 1.1274 & 0.0113\\
    \hline
    SHRED & Arbitrary&69.8\% & 0.6\% & 0.1\% & 0.05\%\hspace{-0.6em} & 1.1522 & 0.0214\\
    BY & Arbitrary & 68.4\% & 0.7\% & 0.9\% & 0.4\% & 1.1449 & 0.0103\\
    \hline
  \end{tabular}
\end{table}

\begin{table}[H]
  \caption{\label{tab09} Gaussian regression, $n = 1000, p = 300, T = 100$, clustered correlation $\rho = 0.3$ to $\rho = 0.6$}
  \centering
  \small
  \begin{tabular}{llccrccc}
    \hline
    \textbf{Method} & \textbf{FDR Cut-off} & \textbf{Power} & \textbf{SE Power} & \textbf{FDR} & \textbf{SE FDR} & \textbf{MSE} & \textbf{SE MSE} \\
    \hline
    Lasso & No FDR cut-off & 84.4\% & 0.7\% & 65.2\% & 0.3\% & 1.3334 & 0.0287\\
    \hline
    Mirror& MDS (Data driven) &61.2\% & 1.0\% & 6.7\% & 0.9\% & 2.1191 & 0.0745\\
    Mirror & DS (Data driven) &20.8\% & 1.2\% & 7.5\% & 1.0\% & 4.2313 & 0.2349\\
    Knockoff & Data driven &  39.9\% & 1.5\% & 12.6\% & 1.2\% & 2.1899 & 0.1075\\
    \hline
    BH & PRDS &74.1\% & 0.5\% & 2.5\% & 0.5\% & 1.5908 & 0.0615 \\
    SHRED & PRDS & 75.2\% & 0.5\% & 1.4\% & 0.5\% & 1.4875 & 0.0724\\ 
    SHRED & Heuristic & 78.3\% & 0.6\% & 3.1\% & 0.5\% & 1.4678 & 0.0828\\
    SHREDDER &PPRDS & 76.2\% & 0.7\% & 2.3\% & 0.4\% & 1.5199 & 0.0915\\
    \hline
    SHRED & Arbitrary &72.7\% & 0.6\% & 0.2\% & 0.5\% & 1.5308 & 0.0610\\
    BY & Arbitrary & 71.8\% & 0.7\% & 0.7\% & 0.1\% & 1.5646 & 0.0871\\
    \hline
  \end{tabular}
\end{table}

\begin{table}[H]
  \caption{\label{tab011} Gaussian regression, $n = 1000, p = 300, T = 100$, clustered correlation correlation $\rho = 0.6$ to $\rho = 0.9$}
  \centering
  \small
  \begin{tabular}{llccrccc}
    \hline
    \textbf{Method} & \textbf{FDR Cut-off} & \textbf{Power} & \textbf{SE Power} & \textbf{FDR} & \textbf{SE FDR} & \textbf{MSE} & \textbf{SE MSE} \\
    \hline
    Lasso& No FDR cut-off & 80.8\% & 0.4\% & 57.7\% & 0.4\% & 1.3087 & 0.0168\\
    \hline
    Mirror & MDS (Data driven) &52.3\% & 1.7\% & 14.3\% & 1.2\% & 3.6486 & 0.0823\\
    Mirror & DS (Data driven) &14.1\% & 0.7\% & 15.0\% & 1.3\% & 6.5350 & 0.1427\\
    Knockoff & Data driven &   39.4\% & 1.3\% & 12.3\% & 1.2\% & 3.3186 & 0.0909\\
    \hline
    BH & PRDS & 68.4\% & 0.7\% & 2.1\% & 0.3\% & 1.6703 & 0.0525 \\
    SHRED & PRDS & 73.9\% & 0.8\% & 2.3\% & 0.4\% & 1.5845 & 0.0848\\ 
    SHRED & Heuristic & 78.1\% & 0.8\% & 3.2\% & 0.5\% & 1.5703 & 0.0892\\
    SHREDDER & PPRDS & 74.8\% & 0.7\% & 2.9\% & 0.5\% & 1.5735 & 0.0738\\
    \hline
    SHRED & Arbitrary &67.8\% & 0.9\% & 0.3\% & 0.05\%\hspace{-0.6em} & 1.6816 & 0.0498\\
    BY & Arbitrary &  65.4\% & 0.9\% & 1.3\% & 0.4\% & 1.7291 & 0.0627\\
    \hline
  \end{tabular}
\end{table}

\begin{table}[H]
  \caption{\label{tab14} Gaussian regression, $n = 1000, p = 300, T = 100$, AR(1) correlation $\rho = 0.6$}
  \centering
  \small
  \begin{tabular}{llccrccc}
    \hline
    \textbf{Method} & \textbf{FDR Cut-off} & \textbf{Power} & \textbf{SE Power} & \textbf{FDR} & \textbf{SE FDR} & \textbf{MSE} & \textbf{SE MSE} \\
    \hline
    Lasso & No FDR cut-off & 96.1\% & 0.4\% & 52.1\% & 0.1\% & 1.2100 & 0.0311\\
    \hline
    Mirror & MDS (Data driven) &85.1\% & 0.3\% & 5.1\% & 0.5\% & 1.2278 & 0.0318\\
    Mirror & DS (Data driven) &82.5\% & 0.6\% & 3.6\% & 0.4\% & 1.2831 & 0.0300\\
    Knockoff & Data driven & 82.4\% & 0.6\% & 3.3\% & 0.5\% & 1.2980 & 0.0414\\
    \hline
    BH & PRDS & 84.7\% & 0.9\% & 3.4\% & 0.6\% & 1.1922 & 0.0223 \\
    SHRED & PRDS & 87.9\% & 0.4\% & 1.6\% & 0.1\% & 1.1583 & 0.0191\\ 
    SHRED & Heuristic & 89.9\% & 0.5\% & 3.2\% & 0.3\% & 1.1498 & 0.0175\\
    SHREDDER & PPRDS & 87.9\% & 0.4\% & 1.8\% & 0.3\% & 1.1815 & 0.0195\\
    \hline
    SHRED & Arbitrary &84.5\% & 0.5\% & 0.4\% & 0.05\%\hspace{-0.6em} & 1.1648 & 0.0571\\
    BY & Arbitrary & 82.4\% & 0.8\% & 1.2\% & 0.3\% & 1.2131 & 0.0591\\
    \hline
  \end{tabular}
\end{table}

\begin{table}[H]
  \caption{\label{tab16} Gaussian regression, $n = 1000, p = 300, T = 100$, AR(1) correlation $\rho = 0.9$}
  \centering
  \small
  \begin{tabular}{llccrccc}
    \hline
    \textbf{Method} & \textbf{FDR Cut-off} & \textbf{Power} & \textbf{SE Power} & \textbf{FDR} & \textbf{SE FDR} & \textbf{MSE} & \textbf{SE MSE} \\
    \hline
    Lasso & No FDR cut-off & 92.7\% & 0.4\% & 51.7\% & 0.4\% & 1.1550 & 0.0170\\
    \hline
    Mirror & MDS (Data driven) &74.3\% & 0.8\% & 5.8\% & 0.5\% & 1.1832 & 0.0287\\
    Mirror & DS (Data driven) & 68.6\% & 1.4\% & 2.5\% & 0.7\% & 1.3509 & 0.0622\\
    Knockoff & Data driven &58.0\% & 1.3\% & 1.1\% & 0.2\% & 2.2691 & 0.0714\\
    \hline
    BH & PRDS & 73.7\% & 0.9\% & 3.7\% & 0.6\% & 1.2407 & 0.0204 \\
    SHRED & PRDS & 74.9\% & 0.6\% & 1.6\% & 0.2\% & 1.2105 & 0.0213\\ 
    SHRED & Heuristic &77.8\% & 0.6\% & 4.3\% & 0.4\% & 1.2017 & 0.0187\\
    SHREDDER & PPRDS &75.5\% & 0.6\% & 2.8\% & 0.3\% & 1.2051 & 0.0117\\
    \hline
    SHRED & Arbitrary &72.6\% & 0.6\% & 0.1\% & 0.05\%\hspace{-0.6em} & 1.2770 & 0.0213\\
    BY & Arbitrary & 69.4\% & 0.7\% & 0.9\% & 0.3\% & 1.4048 & 0.0198\\
    \hline
  \end{tabular}
\end{table}

\subsection{Logistic Regression}

\begin{table}[H]
  \caption{\label{tab18} Logistic regression, $n = 5000, p = 200, T = 100$, single common component correlation $\rho = 0.5$}
  \centering
  \small
  \begin{tabular}{llccrcccc}
    \hline
    \textbf{Method} & \textbf{FDR Cut-off} & \textbf{Power} & \textbf{SE Power} & \textbf{FDR} & \textbf{SE FDR} & \textbf{Accuracy} & \textbf{SE Accuracy} & \textbf{LL} \\
    \hline
    Lasso & No FDR Cut-off & 96.8\% & 0.3\% & 42.9\% & 0.4\% & 92.1562\% & 0.3008\% & -727.7\\
    \hline
    Mirror & MDS (Data driven) & 76.6\% & 0.7\% & 3.0\% & 0.4\% & 90.8892\% & 0.3186\% & -780.1\\
    Mirror & DS (Data driven) & 70.9\% & 0.7\% & 2.9\% & 0.5\% & 90.4663\% & 0.2590\% & -840.5\\
    Knockoff & Data driven & 37.6\% & 0.2\% & 1.4\% & 0.1\% & 87.9016\% &0.1897\% & -893.7\\
    \hline
    BH & PRDS& 76.5\% & 0.6\% & 3.1\% & 0.3\% & 90.6461\% & 0.2985\% & -769.8 \\
    SHRED & PRDS & 77.3\% & 0.6\% & 1.3\% & 0.2\% & 91.1502\% & 0.1916\% & -770.2\\ 
    SHRED & Heuristic & 81.6\% & 0.6\% & 4.7\% & 0.2\% & 91.6991\% & 0.3007\% & -748.3\\
    SHREDDER & PPRDS & 78.1\% & 0.6\% & 3.3\% & 0.2\% & 91.2882\% & 0.1677\% & -757.5 \\
    \hline
    SHRED & Arbitrary & 73.9\% & 0.6\% & 0.2\% & 0.07\%\hspace{-0.6em} & 90.8568\% & 0.1481\% & -788.0\\
    BY & Arbitrary & 73.9\% & 0.7\% & 0.7\% & 0.1\% & 90.6332\% & 0.2845\% & -814.9\\
    \hline
  \end{tabular}
\end{table}	

\begin{table}[H]
  \caption{\label{tab18} Logistic regression, $n = 5000, p = 200, T = 100$, single common component correlation $\rho = 0.9$}
  \centering
  \small
  \begin{tabular}{llccrcccc}
    \hline
    \textbf{Method} & \textbf{FDR Cut-off} & \textbf{Power} & \textbf{SE Power} & \textbf{FDR} & \textbf{SE FDR} & \textbf{Accuracy} & \textbf{SE Accuracy} & \textbf{LL} \\
    \hline
    Lasso & No FDR Cut-off & 88.2\% & 0.7\% & 38.2\% & 0.4\% & 90.9191\% & 0.5907\% & \phantom{1}-927.9\\
    \hline
    Mirror & MDS (Data driven) & 51.6\% & 0.9\% & 3.0\% & 0.5\% & 88.3088\% & 0.7118\% & -1014.4 \\
    Mirror & DS (Data driven) & 46.4\% & 0.9\% & 2.0\% &  0.3\% & 88.0213\% & 0.6993\% & -1023.7\\
    Knockoff & Data driven & 19.5\% & 0.2\% & 0.8\% & 0.1\% & 87.0981\% &0.6033\% & -1175.2\\
    \hline
    BH & PRDS & 52.7\% & 0.6\% & 3.5\% & 0.4\% & 88.8882\% & 0.5762\% & -1009.7\\ 
    SHRED & PRDS & 53.1\% & 0.5\% & 1.2\% & 0.2\% & 88.9727\% & 0.5944\% & -1013.8\\ 
    SHRED & Heuristic & 58.9\% & 0.4\% & 4.5\% & 0.2\% & 89.5810\% & 0.6211\% & -1003.4\\
    SHREDDER & PPRDS & 54.7\% & 0.6\% & 2.9\% & 0.3\% & 89.2822\% & 0.5677\% & -1005.6\\
    \hline
    SHRED & Arbitrary & 49.2\% & 0.6\% & 0.3\% & 0.1\% & 88.5779\% & 0.6115\% & -1048.4\\
    BY & Arbitrary & 48.2\% & 0.7\% & 0.7\% & 0.2\% & 88.5112\% & 0.6869\% & -1045.3\\
    \hline
  \end{tabular}
\end{table}	

\begin{table}[H]
  \caption{\label{tab22} Logistic regression, $n = 5000, p = 200, T = 100$, clustered correlation $\rho = 0.3$ to $\rho = 0.6$}
  \centering
  \small
  \begin{tabular}{llccrcccc}
    \hline
    \textbf{Method} & \textbf{FDR Cut-off} & \textbf{Power} & \textbf{SE Power} & \textbf{FDR} & \textbf{SE FDR} & \textbf{Accuracy} & \textbf{SE Accuracy} & \textbf{LL} \\
    \hline
    Lasso & No FDR Cut-off & 91.5\% & 0.6\% & 39.7\% & 0.4\% & 91.8013\% & 0.2876\% & -611.9 \\
    \hline
    Mirror & MDS (Data driven) & 60.3\% & 0.9\% & 4.7\% & 0.4\% & 90.9192\% & 0.1810\% & -706.4\\
    Mirror & DS (Data driven) & 46.2\% & 0.8\% & 4.8\% & 0.2\% & 88.3909\% & 0.3223\% & -765.3\\
    Knockoff & Data driven & 13.1\%  & 0.8\%  & 0.2\% & 0.1\% & 81.1001\% & 0.5257\% & -935.5\\
    \hline
    BH & PRDS & 61.0\% & 0.6\% & 4.3\% & 0.4\% & 90.8461\% & 0.2223\% & -703.6\\
    SHRED & PRDS & 63.2\% & 0.6\% & 1.4\% & 0.3\% & 91.2971\% & 0.2352\% & -669.3\\ 
    SHRED & Heuristic & 65.8\% & 0.5\% & 4.2\% & 0.4\% & 91.7951\% & 0.1365\% & -617.5\\
    SHREDDER & PPRDS &65.2\% & 0.5\% & 3.4\% & 0.4\% & 91.4110\% & 0.2059\% & -666.3\\
    \hline
    SHRED & Arbitrary &59.9\% & 0.5\% & 0.4\% & 0.03\% & 90.6456\% & 0.1986\% & -712.5 \\
    BY & Arbitrary & 58.3\% & 0.5\% & 0.7\% & 0.3\% & 89.5713\% & 0.3076\% & -725.5\\
    \hline
  \end{tabular}
\end{table}

\begin{table}[H]
  \caption{\label{tab24} Logistic regression, $n = 5000, p = 200, T = 100$, clustered correlation $\rho = 0.6$ to $\rho = 0.9$}
  \centering
  \small
  \begin{tabular}{llccrcccr}
    \hline
    \textbf{Method} & \textbf{FDR Cut-off} & \textbf{Power} & \textbf{SE Power} & \textbf{FDR} & \textbf{SE FDR} & \textbf{Accuracy} & \textbf{SE Accuracy} & \multicolumn{1}{c}{\textbf{LL}} \\
    \hline
    Lasso & No FDR Cut-off & 77.3\% & 0.6\% & 38.8\% & 0.7\% & 91.4874\% & 0.3097\% & -778.1 \\
    \hline
    Mirror & MDS (Data driven) & 51.0\% & 0.6\% & 6.9\% & 0.2\% & 88.3108\% & 0.3244\% & -905.3\\
    Mirror & DS(Data driven) & 20.3\% & 0.6\% & 5.3\% & 0.3\% & 86.4228\% & 0.1798\% & -1086.4\\
    Knockoff & Data driven & 10.5\%  & 1.0\% & 0.01\%\hspace{-0.6em} & 0.01\%\hspace{-0.6em} & 77.5961\% & 0.5015\% & -1231.2\\
    \hline
    BH & PRDS &54.2\% & 0.7\% & 3.8\% & 0.3\% & 88.5001\% & 0.3119\% & -898.3\\
    SHRED & PRDS & 56.8\% & 0.7\% & 1.3\% & 0.3\% & 89.7150\% & 0.1991\% & -893.4\\ 
    SHRED & Heuristic & 60.2\% & 0.6\% & 4.7\% & 0.3\% & 90.0983\% & 0.1892\% & -886.5\\
    SHREDDER & PPRDS &58.8\% & 0.7\% & 1.7\% & 0.4\% & 89.6112\% & 0.3024\% & -891.6\\
    \hline
    SHRED & Arbitrary &51.8\% & 1.1\% & 0.1\% & 0.05\%\hspace{-0.6em} & 88.9037\% & 0.4003\% & -903.1 \\
    BY & Arbitrary & 49.3\% & 0.9\% & 0.6\% & 0.4\% & 87.8916\% & 0.4101\% & -916.5\\
    \hline
  \end{tabular}
\end{table}

\begin{table}[H]
  \caption{\label{tab18} Logistic regression, $n = 5000, p =
    200, T = 100$, AR(1) correlation $\rho = 0.6$} \centering
  \small
  \begin{tabular}{llccrcccc}
    \hline
    \textbf{Method} & \textbf{FDR Cut-off} & \textbf{Power} & \textbf{SE Power} & \textbf{FDR} & \textbf{SE FDR} & \textbf{Accuracy} & \textbf{SE Accuracy} & \textbf{LL} \\
    \hline
    Lasso & No FDR Cut-off & 91.6\% & 0.6\% & 42.6\% & 0.5\% & 91.7359\% & 0.2021\% & -641.6\\
    \hline
    Mirror & MDS (Data driven) & 77.1\% & 0.6\% & 2.0\% & 0.5\% & 91.7118\% & 0.1420\% & -721.7\\
    Mirror & DS (Data driven) & 75.8\% & 0.7\% & 2.1\% & 0.5\% & 91.3000\% & 0.1358\% & -725.8\\
    Knockoff & Data driven & 33.9\% & 0.4\% & 1.3\% & 0.2\% & 89.1149\% & 0.1182\% & -806.2 \\
    \hline
    BH & PRDS & 76.5\% & 0.6\% & 2.2\% & 0.3\% & 91.5112\% & 0.0953\% & -724.2\\
    SHRED & PRDS & 77.0\% & 0.5\% & 1.2\% & 0.2\% & 91.8896\% & 0.0893\% & -728.8\\ 
    SHRED & Heuristic & 80.4\% & 0.5\% & 3.8\% & 0.4\% & 92.1404\% & 0.1355\% & -711.8\\
    SHREDDER & PPRDS & 79.3\% & 0.6\% & 3.7\% & 0.3\% & 91.9986\% & 0.0887\% & -717.3\\
    \hline
    SHRED & Arbitrary & 73.3\% & 0.6\% & 0.2\% & 0.01\% & 91.6456\% & 0.1003\% & -757.9\\
    BY & Arbitrary & 72.5\% & 0.6\% & 0.8\% & 0.3\% & 91.2325\% & 0.1134\% & -758.8\\
    \hline
  \end{tabular}
\end{table}

\begin{table}[H]
  \caption{\label{tab18} Logistic regression, $n = 5000, p = 200, T = 100$,  AR(1) correlation $\rho = 0.9$}
  \centering
  \small
  \begin{tabular}{llccccccc}
    \hline
    \textbf{Method} & \textbf{FDR Cut-off} & \textbf{Power} & \textbf{SE Power} & \textbf{FDR} & \textbf{SE FDR} & \textbf{Accuracy} & \textbf{SE Accuracy} & \textbf{LL} \\
    \hline
    Lasso & No FDR Cut-off & 89.0\% & 0.7\% & 38.7\% & 0.4\% & 91.3000\% & 0.2000\% & -664.4\\
    \hline
    Mirror & MDS (Data driven) & 53.5\% & 0.8\% & 5.5\% & 0.4\% & 90.6000\% & 0.2000\% & -822.6 \\\
    Mirror & DS (Data driven) & 49.7\% & 0.7\% & 5.0\% & 0.5\% & 88.7522\% & 0.2158\% & -834.9\\
    Knockoff & Data driven & 15.7\% & 0.5\% & 0.6\% & 0.1\% & 85.3180\% & 0.1589\% & -1027.6 \\
    \hline
    BH & PRDS & 52.3\% & 0.7\% & 3.3\% & 0.3\% & 91.1372\% & 0.1301\% & -827.8\\
    SHRED & PRDS & 53.9\% & 0.6\% & 1.6\% & 0.2\% & 91.4110\% & 0.1152\% & -819.6\\ 
    SHRED & Heuristic & 57.7\% & 0.6\% & 4.3\% & 0.4\% & 91.6380\% & 0.1355\% & -773.4\\
    SHREDDER & PPRDS & 55.6\% & 0.7\% & 3.7\% & 0.3\% & 91.4732\% & 0.1356\% & -784.9\\
    \hline
    SHRED & Arbitrary & 45.1\% & 0.6\% & 0.3\% & 0.1\% & 88.9065\% & 0.1509\% & -948.3 \\
    BY & Arbitrary & 43.8\% & 0.6\% & 0.9\% & 0.2\% & 88.5072\% & 0.2137\% & -951.3\\
    \hline
  \end{tabular}
\end{table}

\subsection{Poisson Regression}

\begin{table}[H]
  \caption{\label{tabXX} Poisson regression, $n = 5000, p = 200, T = 100$, independent predictors}
  \centering
  \small
  \begin{tabular}{llccrcccc}
    \hline
    \textbf{Method} & \textbf{FDR Cut-off} & \textbf{Power} & \textbf{SE Power} & \textbf{FDR} & \textbf{SE FDR} & \textbf{MSE} & \textbf{SE MSE} & \textbf{LL}\\
    \hline
    Lasso & No FDR cut-off & 89.9\% & 1.2\% & 43.6\% & 0.7\% & 1.5200 & 0.1521 & -1023.4\\
    \hline
    Mirror & MDS (Data driven) & 52.8\% & 1.1\% & 4.5\% & 0.6\% & 1.6845 & 0.1410 & -1280.6\\
    Mirror & DS (Data driven) & 48.7\% & 1.0\% & 3.3\% & 0.6\% & 1.8189 & 0.1528 & -1312.3\\
    \hline
    BH & PRDS & 58.7\% & 0.4\% & 1.6\% & 0.2\% & 1.5690 & 0.1211 & -1222.8\\
    SHRED & PRDS & 58.6\% & 0.6\% & 1.8\% & 0.3\% & 1.5603 & 0.1136 & -1221.9\\
    SHRED & Heuristic & 60.1\% & 0.5\% & 3.1\% & 0.2\% & 1.5249 & 0.1108 &-1220.4\\
    SHREDDER & PPRDS & 58.8\% & 0.5\% & 2.1\% & 0.3\% & 1.5595 & 0.1009 &-1221.7\\
    \hline
    SHRED & Arbitrary & 51.3\% & 0.4\% & 0.6\% & 0.2\% & 1.6808 & 0.1189 &-1253.4\\
    BY & Arbitrary & 51.2\% & 0.3\% & 0.5\% & 0.2\% & 1.6872 & 0.1289 &-1254.0\\
    \hline
  \end{tabular}
\end{table}

\begin{table}[H]
  \caption{\label{tab30} Poisson regression, $n = 5000, p = 200, T = 100$, single common component correlation $\rho = 0.5$}
  \centering
  \small
  \begin{tabular}{llrcrcccr}
    \hline
    \textbf{Method} & \textbf{FDR Cut-off} & \textbf{Power} & \textbf{SE Power} & \textbf{FDR} & \textbf{SE FDR} & \textbf{MSE} & \textbf{SE MSE} & \multicolumn{1}{c}{\textbf{LL}} \\
    \hline
    Lasso & No FDR Cut-off & 88.3\% & 1.4\% & 40.2\% & 0.9\% & 1.5331 & 0.1580 & -1064.3\\
    \hline
    Mirror & MDS (Data driven) & 40.1\% & 1.6\% & 2.1\% & 0.4\% & 1.6742 & 0.1287 & -1519.1\\
    Mirror & DS (Data driven) & 18.4\% & 1.6\% & 0.8\% & 0.4\% & 1.8560 & 0.1519 & -1763.9\\
    Knockoff & Data Driven & 0\% & NA & 0\% & NA & NA & NA & NA \\
    \hline
    BH & PRDS & 55.0\% & 0.4\% & 2.7\% & 0.9\% & 1.6544 & 0.1645 & -934.9\\
    SHRED & PRDS & 56.5\% & 0.7\% & 1.8\% & 0.4\% & 1.5224 & 0.1509 & -913.7\\ 
    SHRED &Heuristic &60.8\% & 0.4\% & 3.0\% & 0.6\% & 1.3308 & 0.0895 & -879.9\\
    SHREDDER & PPRDS & 57.7\% & 0.4\% & 2.3\% & 0.5\% & 1.4196 & 0.0649 & -912.6\\
    \hline
    SHRED & Arbitrary &50.2\% & 0.7\% & 0.1\% & 0.01\%\hspace{-0.6em} & 1.6635 & 0.1267 & -944.9\\
    BY & Arbitrary &49.3\% & 0.8\% & 1.1\% & 0.9\% & 1.6821 & 0.1424 & -1173.6\\
    \hline
  \end{tabular}
\end{table}

\begin{table}[H]
  \caption{\label{tab32} Poisson regression, $n = 5000, p = 200, T = 100$, single common component correlation $\rho = 0.9$}
  \centering
  \small
  \begin{tabular}{llrcrcccc}
    \hline
    \textbf{Method} & \textbf{FDR Cut-off} & \textbf{Power} & \textbf{SE Power} & \textbf{FDR} & \textbf{SE FDR} & \textbf{MSE} & \textbf{SE MSE} & \textbf{LL} \\
    \hline
    Lasso & No FDR Cut-off & 69.3\% & 1.7\% & 37.9\% & 1.1\% & 1.5119 & 0.1215 & -1348.7\\
    \hline
    Mirror & MDS (Data driven) &21.0\% & 0.6\% & 5.8\% & 0.8\% & 1.7272 & 0.1109 & -1531.1\\
    Mirror & DS (Data driven) & 14.0\% & 0.8\% & 4.0\% & 0.5\% & 1.8631 & 0.1213 & -1839.9\\
    Knockoff & Data Driven & 0\% & NA & 0\% & NA & NA & NA & NA \\
    \hline
    BH & PRDS & 38.8\% & 0.7\% & 3.3\% & 1.0\% & 1.5769 & 0.0919 & -1532.6\\
    SHRED & PRDS & 41.8\% & 0.8\% & 1.4\% & 0.1\% & 1.5046 & 0.0819 & -1250.6\\ 
    SHRED & Heuristic & 45.4\% & 0.7\% & 3.2\% & 0.1\% & 1.4161 & 0.0800 & -1250.4\\
    SHREDDER & PPRDS & 42.8\% & 0.3\% & 1.8\% & 0.8\% & 1.4446 & 0.0799 & -1257.5\\
    \hline
    SHRED & Arbitrary& 34.9\% & 0.9\% & 0.2\% & 0.02\%\hspace{-0.6em} & 1.6211 & 0.0831 & -1535.8\\
    BY & Arbitrary & 32.0\% & 0.7\% & 1.3\% & 0.7\% & 1.6659 & 0.1459 & -1552.7\\
    \hline
  \end{tabular}
\end{table}

\begin{table}[H]
  \caption{\label{tab34} Poisson regression, $n = 5000, p = 200, T = 100$, clustered correlation $\rho = 0.3$ to $\rho = 0.6$}
  \centering
  \small
  \begin{tabular}{llrcrcccc}
    \hline
    \textbf{Method} & \textbf{FDR Cut-off} & \textbf{Power} & \textbf{SE Power} & \textbf{FDR} & \textbf{SE FDR} & \textbf{MSE} & \textbf{SE MSE} & \textbf{LL} \\
    \hline
    Lasso & No FDR Cut-off & 76.2\% & 2.4\% & 37.9\% & 1.4\% & 1.6099 & 0.1814 & -1305.8\\
    \hline
    Mirror & MDS (Data driven) & 30.4\% & 1.2\% & 7.5\% & 0.3\% & 2.6833 & 0.1795 & -1989.3\\
    Mirror & DS (Data driven) & 8.9\% & 1.7\% & 6.0\% & 0.6\% & 4.9602 & 0.3427 & -3816.7\\
    Knockoff & Data Driven & 0\% & NA & 0\% & NA & NA & NA & NA \\
    \hline
    BH &PRDS &35.3\% & 1.4\% & 2.5\% & 0.5\% & 2.4304 & 0.2109 & -1798.4\\
    SHRED & PRDS & 36.4\% & 0.6\% & 1.6\% & 0.5\% & 2.3435 & 0.1499 & -1696.6\\ 
    SHRED & Heuristic & 39.3\% & 0.9\% & 1.7\% & 0.4\% & 2.2038 & 0.1916 & -1658.8\\
    SHREDDER & PPRDS & 37.0\% & 0.8\% & 1.2\% & 0.2\% & 2.2548 & 0.1908 & -1687.5\\
    \hline
    SHRED & Arbitrary & 34.1\% & 0.9\% & 0.2\% & 0.03\%\hspace{-0.6em} & 2.5448 & 0.1876 & -1801.8\\
    BY & Arbitrary & 33.9\% & 1.1\% & 0.3\% & 0.01\%\hspace{-0.6em} & 2.5932 & 0.2005 & -1801.9\\
    \hline
  \end{tabular}
\end{table}

\begin{table}[H]
  \caption{\label{tab36} Poisson regression, $n = 5000, p = 200, T = 100$, clustered correlation $\rho = 0.6$ to $\rho = 0.9$}
  \centering
  \small
  \begin{tabular}{llrcrcccc}
    \hline
    \textbf{Method} & \textbf{FDR Cut-off} & \textbf{Power} & \textbf{SE Power} & \textbf{FDR} & \textbf{SE FDR} & \textbf{MSE} & \textbf{SE MSE} & \textbf{LL} \\
    \hline
    Lasso & No FDR Cut-off & 74.8\% & 1.6\% & 39.9\% & 1.0\% & 2.1270 & 0.2229 & -2075.5\\
    \hline
    Mirror & MDS (Data driven) & 28.4\% & 1.7\% & 9.5\% & 0.4\% & 3.4712 & 0.2340 & -2362.3\\
    Mirror & DS (Data driven) & 8.1\% & 2.1\% & 6.0\% & 0.8\% & 5.3871 & 0.3202 & -4353.1\\
    Knockoff & Data Driven & 0\% & NA & 0\% & NA & NA & NA & NA \\
    \hline
    BH & PRDS & 31.3\% & 1.5\% & 1.7\% & 0.5\% & 3.2335 & 0.2483 & -2288.8\\
    SHRED & PRDS & 33.7\% & 1.0\% & 1.5\% & 0.5\% & 3.2223 & 0.2012 & -2252.5\\ 
    SHRED & Heuristic & 34.8\% & 1.4\% & 2.3\% & 0.6\% & 3.2891 & 0.2163 & -2239.6\\
    SHREDDER & PPRDS & 33.4\% & 1.3\% & 1.5\% & 0.5\% & 3.2214 & 0.2016 & -2243.7\\
    \hline
    SHRED & Arbitrary & 30.1\% & 1.4\% & 0.3\% & 0.02\%\hspace{-0.6em} & 3.4893 & 0.2315 & -2324.4\\
    BY & Arbitrary &  29.3\% & 1.4\% & 0.4\% & 0.01\%\hspace{-0.6em} & 3.5064 & 0.2328 & -2359.7\\
    \hline
  \end{tabular}
\end{table}

\begin{table}[H]
  \caption{\label{tab38} Poisson regression, $n = 5000, p = 200, T = 100$, AR(1) correlation $\rho = 0.6$}
  \centering
  \small
  \begin{tabular}{llrcrcccc}
    \hline
    \textbf{Method} & \textbf{FDR Cut-off} & \textbf{Power} & \textbf{SE Power} & \textbf{FDR} & \textbf{SE FDR} & \textbf{MSE} & \textbf{SE MSE} & \textbf{LL} \\
    \hline
    Lasso & No FDR Cut-off & 98.0\% & 0.04\% & 40.9\% & 0.8\% & 1.5649 & 0.1180 & -1012.6\\
    \hline
    Mirror & MDS (Data driven) & 73.1\% & 1.3\% & 3.0\% & 0.2\% & 1.8737 & 0.1699 & -1284.5\\
    Mirror & DS (Data driven) & 66.6\% & 1.6\% & 1.3\% & 0.6\% & 1.9498 & 0.1907 & -1440.3\\
    Knockoff & Data Driven & 0\% & NA & 0\% & NA & NA & NA & NA \\
    \hline
    BH & PRDS &83.5\% & 0.2\% & 2.9\% & 0.6\% & 1.6109 & 0.0835 & -1092.4\\
    SHRED & PRDS & 83.8\% & 0.6\% & 1.5\% & 0.4\% & 1.5097 & 0.0949 & -1073.9\\ 
    SHRED & Heuristic & 86.6\% & 0.5\% & 3.6\% & 0.6\% & 1.5126 & 0.0849 & -1061.3\\
    SHREDDER & PPRDS & 85.3\% & 0.6\% & 2.5\% & 0.6\% & 1.5246 & 0.1218 & -1072.4\\
    \hline
    SHRED & Arbitrary & 82.5\% & 0.8\% & 0.2\% & 0.02\%\hspace{-0.6em} & 1.6357 & 0.0811 & -1096.2\\
    BY & Arbitrary & 81.1\% & 0.6\% & 1.4\% & 0.6\% & 1.6886 & 0.0635 & -1094.7\\
    \hline
  \end{tabular}
\end{table}

\begin{table}[H]
  \caption{\label{tab40} Poisson regression, $n = 5000, p = 200, T =
    100$, AR(1) correlation $\rho = 0.9$} \centering \small
  \begin{tabular}{llrcrcccc}
    \hline
    \textbf{Method} & \textbf{FDR Cut-off} & \textbf{Power} & \textbf{SE Power} & \textbf{FDR} & \textbf{SE FDR} & \textbf{MSE} & \textbf{SE MSE} & \textbf{LL} \\
    \hline
    Lasso & No FDR Cut-off & 87.5\% & 1.8\% & 33.4\% & 2.0\% & 2.0505 & 0.2099 & -1556.2\\
    \hline
    Mirror & MDS (Data driven) & 41.1\% & 1.5\% & 1.9\% & 1.0\% & 2.5636 & 0.1897 & -2219.9\\
    Mirror & DS (Data driven) & 20.3\% & 2.0\% & 0.0\% & 0.0\% & 2.6331 & 0.2193 & -3262.8\\
    Knockoff & Data Driven & 0\% & NA & 0\% & NA & NA & NA & NA \\
    \hline
    BH & PRDS & 75.9\% & 0.3\% & 1.8\% & 0.6\% & 1.9089 & 0.0808 & -1596.7\\
    SHRED & PRDS & 78.7\% & 0.8\% & 1.3\% & 0.6\% & 1.8848 & 0.0811 & -1563.9\\ 
    SHRED & Heuristic & 80.8\% & 1.1\% & 3.5\% & 0.7\% & 1.8792 & 0.1017 & -1550.8\\
    SHREDDER & PPRDS & 79.2\% & 0.9\% & 2.1\% & 0.5\% & 1.9101 & 0.1190 & -1583.5\\
    \hline
    SHRED & Arbitrary & 73.8\% & 0.7\% & 0.2\% & 0.01\%\hspace{-0.6em} & 1.9934 & 0.0889 & -1599.7\\
    BY & Arbitrary & 72.4\% & 0.9\% & 0.6\% & 0.07\%\hspace{-0.6em} & 2.0531 & 0.1295 & -1613.7\\
    \hline
  \end{tabular}
\end{table}

\subsection{Comparing Model Performance of SHRED and the BH Methods}

\begin{table}[H]
  \caption{\label{tab41} One sample T-Test $p$-values for $H_A:
    \mu_{\textrm{MSE}_{\textrm{SHRED}}} <
    \mu_{\textrm{MSE}_{\textrm{BH}}}$ and $H_A:
    \mu_{\textrm{Accuracy}_{\textrm{SHRED}}} >
    \mu_{\textrm{Accuracy}_{\textrm{BH}}}$, or similar hypotheses for
    other comparisons.}
  \centering
  \small
  \begin{tabular}{lccc}
    \hline
    \textbf{Simulation} & \textbf{SHRED PRDS vs. BH} & \textbf{SHREDDER vs. BH} & \textbf{SHRED vs. BY} \\
    \hline
    Gaussian, Independent & 0.98 & 0.98 & 0.98\\
    Gaussian, single common component $\rho = 0.5$ & 0.51 & 0.19 & 0.16\\
    Gaussian, single common component $\rho = 0.9$ & 0.15 & 0.14 &
    \cellcolor{yellow!60}  \hspace{-1.2em}< 0.01 \\
    Gaussian, clustered low & \cellcolor{yellow!60} \hspace{-1.2em}< 0.01 & \cellcolor{yellow!60}    \hspace{-1.2em}< 0.01 & \cellcolor{yellow!60}  \hspace{-1.2em}< 0.01 \\
    Gaussian, clustered high & 0.21 & 0.12 & 0.31 \\
    Gaussian, AR(1) $\rho = 0.6$ & 0.35 & 0.13 & 0.25\\
    Gaussian, AR(1), $\rho = 0.9$ & \cellcolor{yellow!60} \phantom{1\!\!}0.044 &
    \cellcolor{yellow!60} \phantom{1\!\!}0.023 & \cellcolor{yellow!60} \hspace{-1.2em}< 0.01\\
    \hline
    Logistic, Independent & 0.25 & 0.37 & 0.35\\
    Logistic, single common component $\rho = 0.5$ &
    \cellcolor{yellow!60} \phantom{1\!\!}0.038 & \cellcolor{yellow!60} 0.022 & 0.24\\
    Logistic, single common component $\rho = 0.9$ & 0.41 & 0.37 & 0.47\\
    Logistic, clustered low & 0.18 & 0.08 & \cellcolor{yellow!60} \hspace{-1.2em}< 0.01 \\
    Logistic, clustered high & \cellcolor{yellow!60} \phantom{1\!\!}0.031 & 0.08 & \cellcolor{yellow!60} \phantom{1\!\!}0.039\\
    Logistic, AR(1) $\rho = 0.6$ & \cellcolor{yellow!60} \hspace{-1.2em}< 0.01 &
    \cellcolor{yellow!60} \hspace{-1.2em}< 0.01 & \cellcolor{yellow!60} \hspace{-1.2em}< 0.01\\
    Logistic, AR(1), $\rho = 0.9$ & \cellcolor{yellow!60} \hspace{-1.2em}< 0.01 & \cellcolor{yellow!60} \hspace{-1.2em}< 0.01 &
    \cellcolor{yellow!60} \hspace{-1.2em}< 0.01\\
    \hline
    Poisson, Independent & 0.48 & 0.47 & 0.48\\
    Poisson, single common component $\rho = 0.5$ & 0.28 & \cellcolor{yellow!60} \phantom{1\!\!}0.043 & 0.58\\
    Poisson, single common component $\rho = 0.9$ & 0.38 & 0.14 & 0.38\\
    Poisson, clustered low & 0.39  & 0.26 & 0.35 \\
    Poisson, clustered high & 0.59 & 0.59 & 0.43 \\
    Poisson, AR(1) $\rho = 0.6$ & 0.30 & 0.27 & 0.16\\
    Poisson, AR(1), $\rho = 0.9$ & 0.32 &  0.57 & 0.46\\
  \end{tabular}
\end{table}

\bibliography{SHRED_reference_list2.bib}